\documentclass[fleqn,useAMS,usenatbib]{mnras}
\usepackage{threeparttable}
\usepackage{amssymb}
\usepackage{bm}
\usepackage{graphicx}
\usepackage{CJKutf8}
\usepackage[T1]{fontenc}
\usepackage{ae,aecompl}

\title[The dependence of metallicity and gas content on environment]{The Dependence of the Mass-Metallicity Relation on Large Scale Environment}

\author[P.-F. Wu et al.]
{Po-Feng Wu (吳柏鋒)$^1$, 
H. Jabran Zahid$^2$,
Ho Seong Hwang$^3$,
Margaret J. Geller$^2$
\\
$^1$Max-Planck Institut f\"{u}r Astronomie, K\"{o}nigstuhl 17, D-69117, Heidelberg, Germany\\
$^2$Smithsonian Astrophysical Observatory, Harvard-Smithsonian Center for Astrophysics - 60 Garden Street, Cambridge, MA 02138\\
$^3$Quantum Universe Center, Korea Institute for Advanced Study, 85 Hoegiro, Dongdaemun-gu, Seoul 02455, Republic of Korea
}

\defcitealias{zah14b}{Z14}
\begin{document}
\begin{CJK*}{UTF8}{bkai}
	
	\maketitle
	
	\begin{abstract}
		We examine the relation between gas-phase oxygen abundance and stellar mass---the MZ relation---as a function of the large scale galaxy environment parameterized by the local density. The dependence of the MZ relation on the environment is small. The metallicity where the MZ relation saturates and the slope of the MZ relation are both independent of the local density. The impact of the large scale environment is completely parameterized by the anti-correlation between local density and the turnover stellar mass where the MZ relation begins to saturate. Analytical modeling suggests that the anti-correlation between the local density and turnover stellar mass is a consequence of a variation in the gas content of star-forming galaxies. Across $\sim1$ order of magnitude in local density, the gas content at a fixed stellar mass varies by $\sim5\%$. Variation of the specific star formation rate with environment is consistent with this interpretation. At a fixed stellar mass, galaxies in low density environments have lower metallicities because they are slightly more gas-rich than galaxies in high density environments. Modeling the shape of the mass-metallicity relation thus provides an indirect means to probe subtle variations in the gas content of star-forming galaxies.

	\end{abstract}
	

\section{Introduction}

Chemical enrichment is important for understanding the evolution of galaxies. Heavy elements, i.e., metals, are synthesized in stars and then released into the interstellar medium (ISM) by stellar winds and supernova explosions. These metals mix with inflowing gas from the circum- and intergalactic medium. The metal-enriched ISM subsequently acts as the raw material for the next generation of stars. The metallicity thus reflects the history of star-formation and gas flows in galaxies. 

The correlation between the stellar mass and gas-phase oxygen abundance is referred to as the mass-metallicity (MZ) relation. \citet{leq79} first showed that galaxies with larger stellar masses have higher metallicities. To first order, the origin of the MZ relation can be understood as a simple process of recycling metals from massive stars into the ISM. Galaxies with larger stellar masses have produced more massive stars and therefore have synthesized and released more heavy elements throughout their lives. Thus, a higher metallicity is naturally expected. 
However, galaxies do not evolve as closed systems. The metal content in the ISM may be altered by inflowing gas from the intergalactic medium and by outflows driven by stellar winds or supernovae. Thus, the metallicity in star-forming galaxies depends on both star-formation and gas flows.  

Analytical chemical evolution models reproduce the average MZ relation \citep{spi10,pee11,lil13,zah14b,spi15}. These models parametrize physical properties of galaxies, including the magnitude and metallicity of inflows and outflows and the star-formation rate (SFR). The combination of these models and measurements of the MZ relation provides constraints for the physical properties of galaxies \citep[e.g.,][]{lil13,pen14,zah14b}. 

Galaxy environment is a potentially important driver of galaxy evolution. Several physical properties of galaxies are correlated with environment, including SFRs, colours, and morphologies \citep{oem74,dre80,has98,kau04,hog04}. The metallicity also appears to be dependent on the environment. In the nearby universe, galaxies in higher local density regions or in clusters have higher metallicities at given stellar mass on average \citep{mou07,coo08,ell09,pet12,pen14}. However, the impact of the environment on metallicity is weak and depends on stellar mass. Low mass galaxies show metallicity vairations of $\lesssim$0.1~dex; in massive galaxies the metallicities are independent of environment. At higher redshifts, the dependence of metallicity on environment is not clear. \citet{kul13} and \citet{shi15} both report that proto-cluster galaxies at $z\sim2$ exhibit metallicity enhancements compared to field galaxies at those redshifts; \citet{val15} report opposite trends. The samples at high redshifts are small and may be subject to selection effects and systematic measurement uncertainties. 

Several mechanisms may explain the dependence of metallicity on environment. The intergalactic medium (IGM) is enriched by metals expelled from galaxies through supernovae and/or stellar winds \citep{opp06,opp08}. The local gas reservoir may be more metal-rich in denser environments. Reaccretion of enriched gas in dense environments may boost the metallicity over galaxies in less dense regions \citep{opp06,pen14}. On the other hand, galaxy mergers, tidal interactions, and ram-pressure stripping can disturb the ISM thus altering properties of the gas reservoir \citep{kew10,tor12}. The effect of these environment dependent mechanisms on the metallicity is poorly constrained.

We investigate the impact of environment on the metallicity using the Universal Metallicity Relation (UMR) formulation of \citet[Z14 hereafter]{zah14b}. Regardless of the detailed properties of gas flows, the impact of the environment on the MZ relation can be well parameterized by a variation in the stellar mass where the MZ relation saturates. This result is a consequence of a variation in the gas content in star-forming galaxies with large scale environment. The environmental dependance of the SFRs of galaxies tests this interpretation. The shape of MZ relation provides a proxy for probing the average gas content in star-forming galaxies. This method is useful beyond the local Universe where direct measurements of ISM gas content are rare.

The data and the derivation of physical properties of galaxies are in Section~\ref{sec:data}. We introduce the chemical evolution model in Section~\ref{sec:mz}. In Section 4 we quantify the impact of the environment on the MZ relation and interpret the origin of this effect physically. We discuss and summarize our results in Section~\ref{sec:dis} and Section~\ref{sec:sum}, respectively.
We adopt the standard cosmology $(H_0, \Omega_m, \Omega_\Lambda) = (70\; \mbox{km s}^{-1}\; \mbox{Mpc}^{-1}, 0.3, 0.7)$ and a \citet{cha03} IMF. 

\section{Data and Analysis}
\label{sec:data}
\begin{figure*}
	\centering
	\includegraphics[width=2\columnwidth]{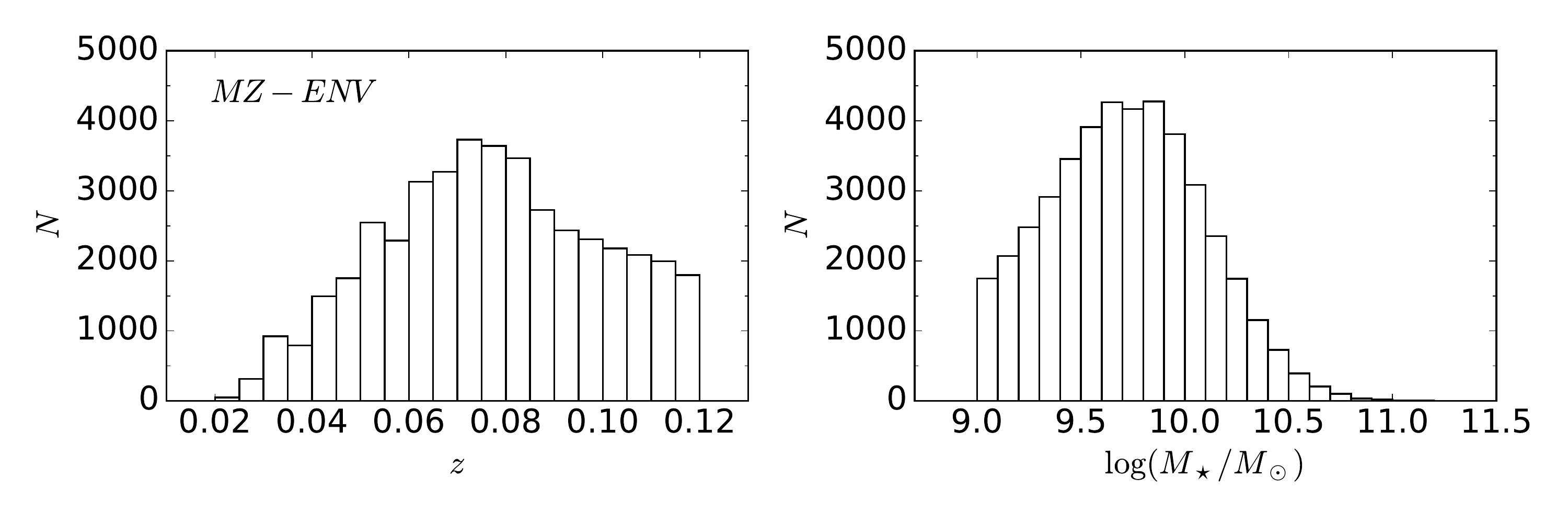}
	\caption{Properties of the MZ-ENV sample. The left and right panels show the distributions in redshift and stellar mass, respectively.}
	\label{fig:prop}
\end{figure*}

We analyze the MZ relation at $z \simeq 0.08$ using data from the SDSS Data Release 7 \citep{aba09}. The sample is similar to the sample analyzed in \citet{zah13} and \citetalias{zah14b}. Here we further select galaxies with local density estimates (see section~\ref{sec:env}). The redshift and stellar mass distributions for the sample analyzed here are shown in Figure~\ref{fig:prop} and we briefly describe the sample below. 

\subsection{The Data}

The SDSS spectroscopic data consist of $\sim 900,000$ galaxies to a limiting magnitude of $r=17.8$, primarily in the redshift range of $0 < z < 0.3$. We adopt the line fluxes measured by the MPA/JHU and the $ugriz$-band c-model magnitudes. We require a signal-to-noise ratio (S/N) $> 3$ in the [O {\sc ii}]$\lambda 3727$ and H$\beta$ emission lines, but apply no S/N cut on the [O {\sc iii}]$\lambda 5007$ line. S/N cuts on the [O {\sc iii}]$\lambda 5007$ emission line are known to bias measurements of the MZ relation at high metallicities \citep{fos12}. We apply the same S/N criteria to the H$\alpha$ and [N {\sc ii}] emission lines in selecting star-forming galaxies. These lines are required to remove AGN. 

We correct for dust extinction in the emission lines by inferring a reddening correction from the Balmer decrement. We derive the intrinsic color excess by assuming an intrinsic H$\alpha$/H$\beta$ ratio of 2.86 \citep{ost89}. We then correct for the dust attenuation using the extinction law of \citet*{car89} and a corresponding $R_v = 3.1$. 

We select galaxies with $z<0.12$ and require a $g-$band fiber aperture covering fraction of $\gtrsim 20\%$ from the comparison of the 3-arcsecond fiber flux with the total flux. This minimum covering fraction ensures the metallicity is representative of the global value \citep{kew05}. We select galaxies with $\log(M_\star/M_\odot) \gtrsim 9.0$ to mitigate incompleteness and ensure reliable determination of metallicity (Section~\ref{sec:r23}). Active galactic nuclei (AGNs) are removed from the sample using the \citet{kew06} classification based on the [N\,{\sc ii}]/H$\alpha$ vs. [O\,{\sc iii}]/H$\beta$ diagram \citep*{bal81}.
We further require that galaxies have estimates of environment and are not near the edge of the survey field (Section~\ref{sec:env}). The final ``MZ-ENV" sample contains 38,434 star-forming galaxies.  

\subsection{Stellar Mass}

We derive stellar masses using the \textit{Le Phare} code \citep{arn99,ilb06}. The stellar masses are determined by $\chi^2$ fitting the observed photometry. The spectral energy distribution (SED) templates are generated with the stellar population synthesis package developed by \citet{bc03}. We adopt an exponentially decreasing star formation history (SFH), SFH $\propto e^{-t/\tau}$, with the time scale $\tau = 0.1, 0.3, 1, 2, 3, 5, 10, 15$, and 30~Gyr. Three different metallicities, $0.2Z_{\odot}$, $0.4Z_{\odot}$ and $Z_{\odot}$ are used. We use a \citet{cal00} extinction law  with $E(B-V)$ = 0, 0.1, 0.2, 0.3, 0.4, 0.5 and 0.6. The root-mean-square uncertainty in stellar mass is $\sim 0.1$ dex \citepalias{zah14b}.

\subsection{Star Formation Rate}

The SFRs for the MZ-ENV sample are derived by the MPA/JHU group using the technique of \citet{bri04} and \citet{sal07}. The strong emission lines of each galaxy are fit using the nebular emission models of \citet{cha01} and all nebular emission lines contribute to the determination of extinction and SFR. Extinction and SFR come primarily from the H$\alpha$/H$\beta$ flux ratio and the H$\alpha$ luminosity, respectively. Details of the technique are in \citet{bri04} and \citet{sal07}. We convert the SFR from a Kroupa to Chabrier IMF by dividing by 1.06. We rely on the relative accuracy in SFR measurements; uncertainties associated with the IMF have no impact on our conclusions.

\subsection{Gas-Phase Oxygen Abundance}
\label{sec:r23}

We derive metallicities using the $R23$ strong line method calibrated by \citet{kk04}. We calculate the relevant ratios of emission fluxes
\begin{equation}
R_{23} = \frac{\mbox{[O\,{\sc ii}]}\lambda 3727 + \mbox{[O\,{\sc iii}]}\lambda 4959 + \mbox{[O\,{\sc iii}]}\lambda 5007}{\mbox{H}\beta}
\end{equation}
and
\begin{equation}
O_{32} = \frac{\mbox{[O\,{\sc iii}]}\lambda 4959 + \mbox{[O\,{\sc iii}]}\lambda 5007}{\mbox{[O\,{\sc ii}]}\lambda 3727}.
\end{equation}
We assume that the flux ratio of [O\,{\sc iii}]$\lambda$5007 to [O\,{\sc iii}]$\lambda$4959 is 3 \citep{ost89} and adopt a value of 1.33 times the [O\,{\sc iii}]$\lambda$5007 intensity when summing the [O\,{\sc iii}]$\lambda$4959 and [O\,{\sc iii}]$\lambda$5007 line fluxes. 

The $R23$ method has two metallicity branches. Because of our mass selection, nearly all of the galaxies in our sample are on the upper branch of the $R23$ diagnostic. We use the [N\,{\sc ii}]$\lambda$6584/H$\alpha$ and [N\,{\sc ii}]$\lambda$6584/[O\,{\sc ii}]$\lambda$3727 line ratios to verify that galaxies in our sample are on the upper branch. Only 47 galaxies ($\sim 0.1\%$ of the sample) are potentially on the lower branch due to either low [N\,{\sc ii}]/[O\,{\sc ii}] or [N\,{\sc ii}]/H$\alpha$ line ratio.

The calibration of \citet{kk04} is based on stellar evolution and photoionization model grids. We repeat our analysis using the empirically calibrated O3N2 metallicity diagnostic \citep{all79} to test whether our results are sensitive to the metallicity diagnostic. The method considers two flux ratios, $O3N2 \equiv \log\{(\mbox{[O\,{\sc iii}]}\lambda5007/\mbox{H}\beta)/(\mbox{[N\,{\sc ii}]}\lambda6583/\mbox{H}\alpha ) \}$, to derive the metallicity. We adopt the calibration of \citet{pet04}, which is anchored on electron temperature-based metallicities from H\,{\sc ii} regions. Our conclusions do not depend on the diagnostic chosen (see Appendix~\ref{app}). 

The intrinsic uncertainty in an individual measurement is $\sim 0.1$~dex \citep{kk04}. We rely only on the relative accuracies of metallicities determined from strong emission lines. Our analysis is not impacted by the uncertainties in the absolute calibration of metallicities \citep{kew08}. Throughout this work, the metallicity is given as a ratio of the number of oxygen atoms to hydrogen atoms and is quoted as $12+\log(O/H)$.

\subsection{Environmental Estimate}
\label{sec:env}

We estimate the local density by calculating the kernel-smoothed $r$-band luminosity density field, following the procedure outlined in \citet{tem12}. The galaxy catalog used for constructing the density field is drawn from the SDSS DR8 \citep{aih11}. \citet{tem12} select galaxies in the main contiguous region of the survey with $r$-band galactic extinction corrected magnitudes $12.5 < m_r < 17.77$. There are 576,493 galaxies in the catalog. All galaxies are $K$ and evolution corrected to $z=0$. The $K$-correction is calculated using the KCORRECT algorithm of \citet{bla07} and the evolution correction is calculated from the luminosity evolution model of \citet{bla03}. Large redshift space distortions due to galaxy peculiar velocities (finger-of-god effect) are suppressed \citep[for details, see][]{tem12}. For the kernel scale adopted in this study, this correction is irrelevant. 

To estimate the density field, the luminosity of each galaxy is first weighted by a distance-dependent factor to take into account the magnitude limit of the survey. \citet{tem12} select galaxies with $12.5 < m_r < 17.77$ for constructing the density field. To account for contribution to the total luminosity by galaxies outside the observed range, \citet{tem12} calculate a weighting factor for each galaxy $W_{d,j}$. This factor is the ratio between the total luminosity of all galaxies relative to the sum of luminosities for galaxies within the luminosity range set by the magnitude limit:
\begin{equation}
W_{d,j} = \frac{\int^{\infty}_{0}Ln(L)\mathrm{d}L}{\int^{L_2}_{L_1}Ln(L)\mathrm{d}L}.
\end{equation}
$L_1$ and $L_2$ are the lower and upper luminosity limits at distance $d$ corresponding to the bright and faint magnitude limits of $m_r = 12.5$ and $m_r = 17.77$, respectively. The average $K$ and evolution correction at each distance are applied to the limits. 
Here $n(L)$ is the luminosity function in the $r$-band determined in the survey region given by \citet{tem12}:
\begin{equation}
n(L)\mathrm{d}L \propto (L/L^\ast)^{-1.305} [1+(L/L^\ast)^{1.81}]^{-3.218} \mathrm{d}(L/L^\ast),
\end{equation}
where the characteristic luminosity $L^\ast$ corresponds to the $r$-band absolution magnitude $M_r = -21.75$.

The density at position $i$ is the sum of the weighted, kernel-smoothed luminosities of all galaxies:
\begin{equation}
\rho_i = \frac{1}{a^3} \sum_{j = 1}^{N} B_3\left(\frac{|\boldsymbol{r}_{j}-\boldsymbol{r}_i|}{a}\right)L_{j} W_{d,j}.
\end{equation}
$L_{j}$ is the luminosity of each galaxy, $a$ is the kernel scale, and $|\boldsymbol{r}_{j} - \boldsymbol{r}_i|$ is the distance between all galaxy at position $j$ and the galaxy at position $i$. The sum is over all $N$ galaxies in the sample. The $B_3$ spline kernel is defined as:
\begin{equation}
B_3(x) = \frac{|x-2|^3 - 4|x-1|^3 + 6|x|^3 - 4|x+1|^3 + |x+2|^3}{12}.
\end{equation}
When $ |\boldsymbol{r}_{j} - \boldsymbol{r}_i|> 2a$, $B_3(x) = 0$. We adopt a kernel scale of 8~Mpc, corresponding to the galaxy correlation length \citep{del88}. 

The luminosity of the galaxy contributes to the density field at its own location. 
To mitigate the self-contribution to the density field, we calculate the local density around each MZ-ENV galaxy at a randomly chosen point within 1~Mpc of the positions of the target galaxy. To minimize the edge effects, we limit our analysis to galaxies that are farther than 8~Mpc from the survey edge. In this paper, we use the \textit{relative} density, the density normalized to the median value, $\log(\rho/\rho_{med})$. 

To test the results, we also derive estimates of the environment using the nearest-neighbor analysis. We calculate both the 3rd and 5th projected nearest-neighbor density and find that our main conclusion (Section~\ref{sec:ori_env}) does not depend on the approach to measuring environment. 


\section{The MZ Relation in the Context of the Universal Metallicity Relation}
\label{sec:mz}

\subsection{Fitting the MZ Relation}

\begin{figure}
	\includegraphics[width=0.95\columnwidth]{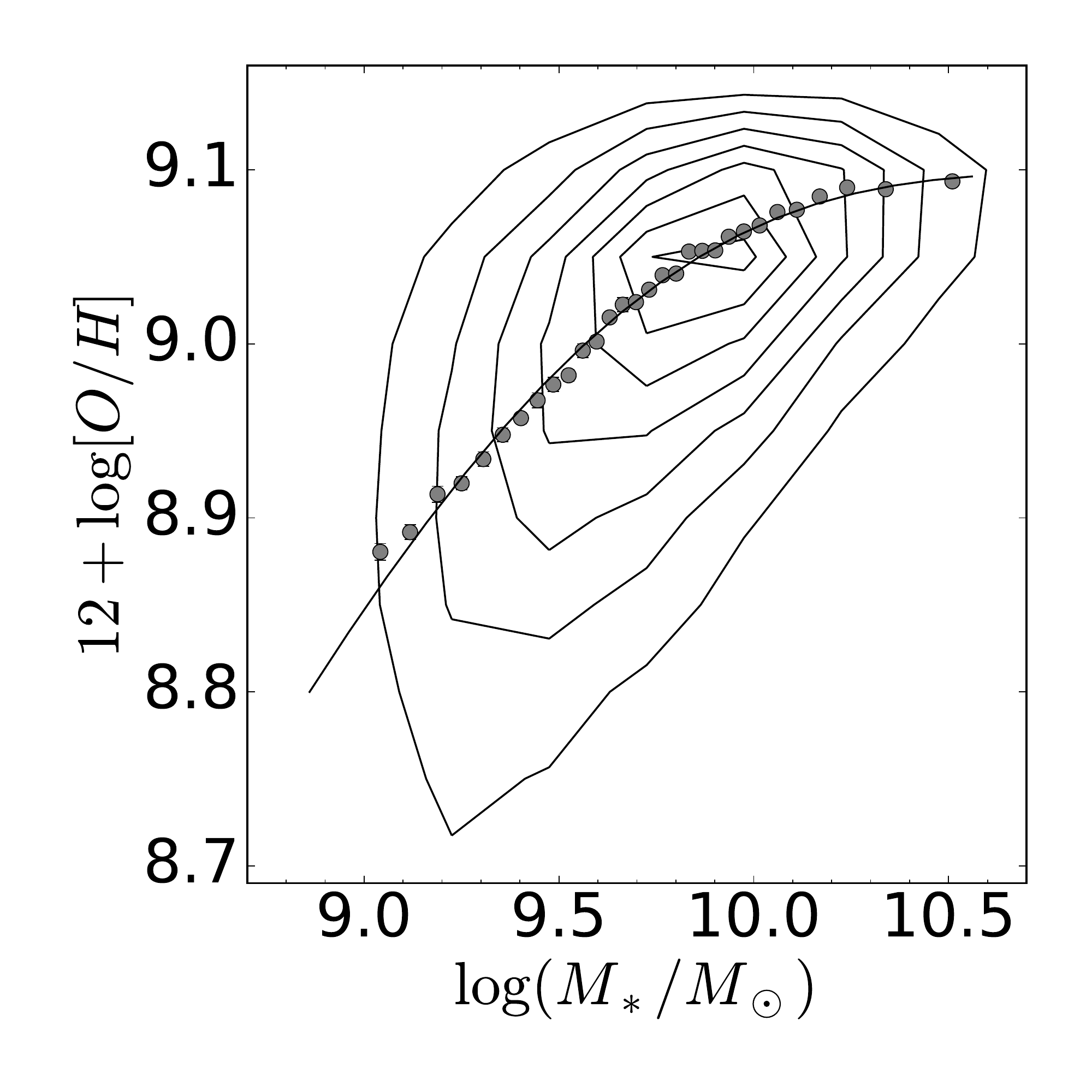}
	\caption{The MZ relation of the MZ-ENV sample. The circles are the median metallicity in each stellar mass bins. The statistical uncertainties in each mass bin is comparable to the size of the circles therefore are not shown here. The solid line is the best-fit model (see text). The contours show the density of the MZ-ENV sample on the stellar mass-metallicity plane.}
	\label{fig:mz}
\end{figure}

Figure~\ref{fig:mz} shows the MZ relation of the MZ-ENV sample. We sort galaxies into equally populated bins in stellar mass and plot the median stellar mass and metallicity for each bin. We determine the statistical uncertainties in the metallicity from bootstrap resampling. The uncertainty is smaller than the symbol size in Figure~\ref{fig:mz}.

We model the MZ relation using the UMR formulation of \citetalias{zah14b}:
\begin{equation}
\label{eq:mz}
12 + \log(O/H) = Z_0 + \log\left[1 - \exp \left( -\left[ \frac{M_\star}{M_0}\right]^\gamma \right) \right].
\end{equation}
The MZ relation is described by three free parameters. $Z_0$ is the saturation metallicity which quantifies the asymptotic upper metallicity limit observed in massive galaxies. $M_0$ is the characteristic mass above which the metallicity asymptotically turns over and approaches the upper metallicity limit, $Z_0$. $\gamma$ is the slope of the MZ relation at the low-mass end where $M_\star \ll M_0$. 

We fit the data using the \textit{MPFIT} package implemented in \textit{IDL} \citep{mar09} and weight the data by the inverse variance determined from bootstrapping. The best-fit parameters and the formal uncertainties of the fit are in Table~\ref{tab:sdss_mz_fit}. To verify the formal uncertainty of the fit given by the \textit{MPFIT} package, we generate 1,000 bootstrap samples and repeat the fitting process. From the distribution of the 1,000 sets of best-fit parameters, we find a good agreement with Table~\ref{tab:sdss_mz_fit}.

\begin{table}
	\caption{Best-fit model parameters for the MZ-ENV galaxies.}
	\label{tab:sdss_mz_fit}
	
	\begin{tabular}{cccc}
		\hline
		\hline
		$\log(\rho/\rho_{med})$ & $Z_0$ & $\log (M_0/M_\odot)$ & $\gamma$ \\ 
		\hline
		Avg.  & $9.100\pm0.001$ & $9.169\pm0.006$ & $0.505\pm0.008$ \\
		Z14 &  $9.102\pm0.002$ & $9.219\pm0.007$ & $0.513\pm0.009$ \\
		\hline
		-0.48 & $9.105\pm0.004$ & $9.212\pm0.016$ & $0.507\pm0.021$ \\ 
		-0.19 & $9.103\pm0.004$ & $9.197\pm0.015$ & $0.496\pm0.019$ \\ 
		0.00 & $9.107\pm0.004$ & $9.174\pm0.016$ & $0.464\pm0.019$ \\ 
		0.18 & $9.102\pm0.004$ & $9.174\pm0.014$ & $0.504\pm0.021$ \\ 
		0.43 & $9.098\pm0.003$ & $9.124\pm0.013$ & $0.498\pm0.018$ \\ 
		\hline
		
	\end{tabular} 
	
\end{table}

The best-fit $Z_o$ and $\gamma$ for the MZ-ENV sample agree with the parameters measured by \citetalias{zah14b} (see Table~\ref{tab:sdss_mz_fit}). The small difference in $M_0$ is a result of sample selection. \citet{tem12} apply strict magnitude selection with $K$ and evolution corrections to the data. At higher redshifts, the \citet{tem12} selection preferentially removes lower mass, high SFR and lower metallicity galaxies in the \citet{zah14b} sample. The exclusion of these galaxies results in the small difference in $M_0$. Here we only measure the relative change in metallicity within a consistently selected sample of galaxies and thus are not affected by the small differences due to selection effects. 

\subsection{Physical Interpretation}

The model we fit to the data (Equation \ref{eq:mz}) is motivated by an analytical solution to the differential equation of galactic chemical evolution. The UMR formulation is based on the inflow model first introduced by \citet{lar72} but with important modifications to account for the impact of gas flows on metallicity. Z14 interpret the origin and evolution of the MZ relation using the UMR formulation.

Z14 find that the metallicity where the MZ relation saturates and the slope of the MZ relation are independent of redshift. The redshift evolution of the MZ relation is solely parameterized by evolution in the turnover mass where the MZ relation begins to saturate. Based on the UMR formulation, the turnover mass is set by the zero-point of the relation between the stellar and gas mass. Thus, the redshift evolution of the MZ relation is a consequence of variations in the gas content of galaxies. We show that this interpretation can explain the dependence of the MZ relation on environment. Details of the UMR formulation can be found in \citetalias{zah14b}, below we discuss the salient features of the model and its derivation.

\citetalias{zah14b} show the change in the gas-phase metallicity with respect to stellar mass can be expressed as 
\begin{equation}
\label{eq:dz}
\frac{dZ}{dM_\star} \approx \frac{Y_N - Z(1-R)}{M_g}.
\end{equation}
We take the metallicity as the mass density of oxygen relative to hydrogen, $Z \equiv M_z/M_g$.  Here, $M_z$ is the mass of oxygen in the gas-phase and $M_g$ is the hydrogen gas mass\footnote{The mass density and number density of oxygen relative to hydrogen in the ISM is related by a constant scaling factor. }. The solution to Equation~\ref{eq:dz} is the metallicity as a function of stellar mass, i.e. the MZ relation.

The numerator of the right-hand-side (RHS) of Equation~\ref{eq:dz} is the net gain of metals in the ISM due to star formation. The first term of the RHS is defined as $Y_N \equiv Y - \zeta$. $Y$ is the nucleosynthetic yield, the mass of oxygen created per unit SFR and $\zeta$ is the amount of oxygen loss due to gas flows per unit SFR. $Y$ does not depend strongly on any galaxy properties \citep{tho98,kob06}, therefore, we treat $Y$ as a constant. For $\zeta$, the detailed process of inflows and outflows and their chemical properties are poorly understood and it is not possible to quantify their impact on metallicity from theoretical arguments alone \citep{zah14a}. Based on multi-epoch observations of the MZ relation and the relation between stellar mass and SFR, \citet{zah12} develop a semi-analytical framework to show that the total mass of oxygen expelled from galaxies over their lifetime is nearly proportional to their stellar mass. This empirical constraint implies that $\zeta$ is approximately constant. Therefore, \citetalias{zah14b} combined $Y$ and $\zeta$ into one constant, $Y_N$, which they refer to as the net yield. Thus, $Y_N$ is the mass of oxygen produced by star formation but accounting for the net amount of oxygen expelled from the ISM. $R$ is the fraction of mass returned to the ISM and thus the second term on the RHS, $Z(1-R)$, represents oxygen that is locked-up in low mass stars forever.

The gas mass of star-forming galaxies is reasonably well described by a power law over approximately four orders of magnitudes in stellar mass \citep[e.g.,][]{pap12}. We parameterize the relation between gas and stellar mass by 
\begin{equation}
\label{eq:gas}
M_g = G M_\star^g, 
\end{equation}
where $G$ is the normalization and $g$ is the power law index. Adopting this relation, the solution to Equation~\ref{eq:dz} is
\begin{equation}
\label{eq:ana}
Z(M_\star) = \frac{Y_N}{1-R}\left[ 1- \exp \left( - \frac{M_\star}{M_g}  \right) \right].
\end{equation}
Here we have dropped a factor (1-R)/(1-g) in the exponent because $R$ and $g$ are nearly equal \citepalias{zah14b}. 

The UMR formulation of \citetalias{zah14b} is based on the direct correspondence between Equations \ref{eq:mz} and \ref{eq:ana}. By taking the logarithm of Equation~\ref{eq:ana}, we can directly relate the parameters we fit to the MZ relation to physical parameters which analytically describe the chemical evolution of galaxies. Thus, the saturation metallicity, $Z_0$, is determined by the net yield and return fraction:
\begin{equation}
\label{eq:z0}
Z_0 = \log\left( \frac{Y_N}{1-R} \right).
\end{equation}
The impact of inflows and outflows on metallicity is encapsulated in $Z_0$. $M_0$ and $\gamma$ together represent the stellar-to-gas mass ratio:
\begin{equation}
\label{eq:m0mg}
\left(\frac{M_\star}{M_0}\right)^\gamma = \frac{M_\star}{M_g}.
\end{equation}
Replacing $M_g$ on the RHS with the relation between gas and stellar mass given in Equation~\ref{eq:gas}, we have
\begin{equation}
\label{eq:gamma}
\gamma = 1-g
\end{equation}
and 
\begin{equation}
\label{eq:m0}
M_0 = G^{1/\gamma}
\end{equation}
In the UMR interpretation, the low-mass-end slope $\gamma$ which we measure by fitting the MZ relation reflects the slope of the relation between gas and stellar mass. The characteristic turnover mass $M_0$ is related to the normalization of the gas and stellar mass relation. The three model parameters we fit to the MZ relation (Equation \ref{eq:mz}) combined with the UMR formulation provide constraints on the physical properties of star-forming galaxies. 

\begin{figure*}
	\centering
	\includegraphics[width=\textwidth]{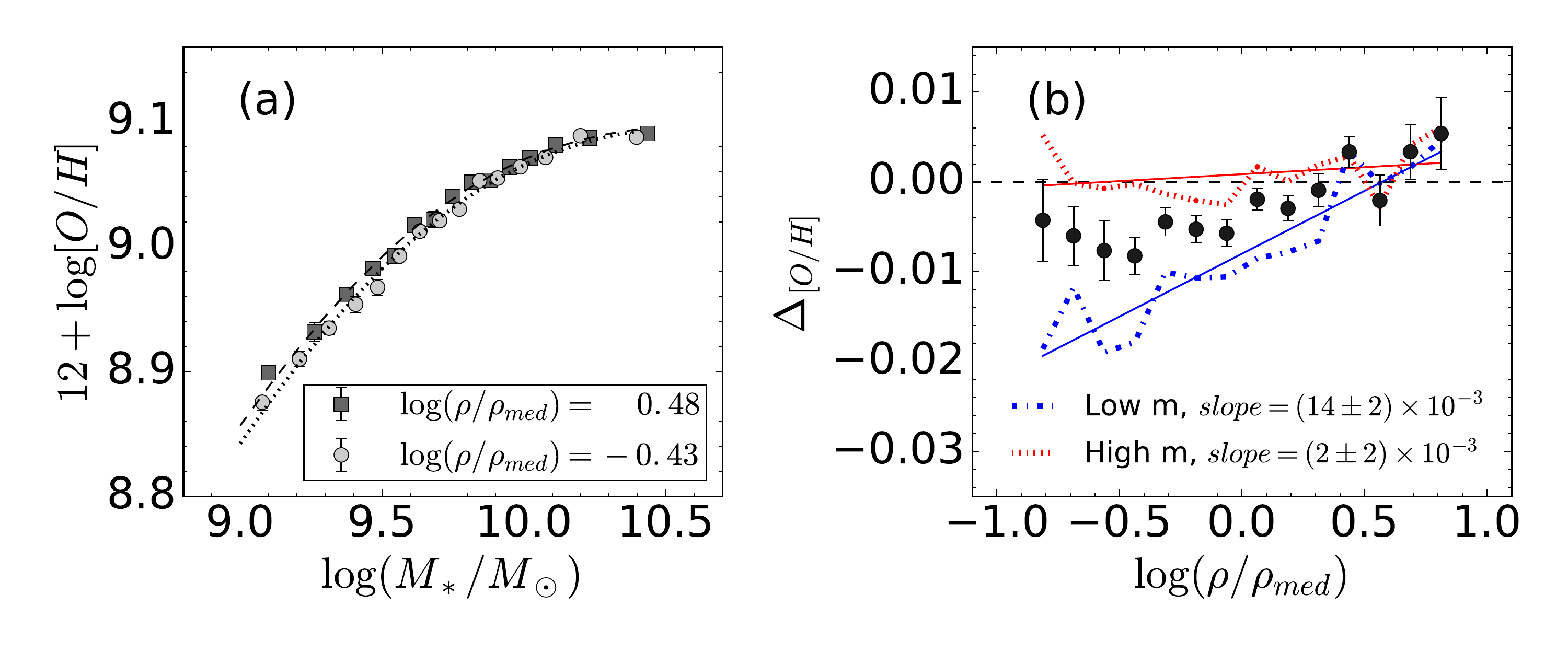}
	\caption{(a) The MZ relations of galaxies at two different local densities. Galaxies at higher densities have higher metallicities at a fixed stellar mass. The dashed (dotted) line is the best-fit model for the high (low) density sample. (b) The metallicity difference between a galaxy and the best-fit MZ relation at the stellar mass of the galaxy, $\Delta_{[O/H]}$, plotted as a function of local density for the MZ-ENV sample. We show the median value in each density bin. The uncertainty is determined from bootstrap resampling. The red dotted line and the blue dash-dotted line are the median $\Delta_{[O/H]}$ for high ($M_\star \geq 10^{9.8} M_\odot$) and low ($M_\star < 10^{9.8} M_\odot$) mass galaxies, respectively. Solid lines are the best-fit $\Delta_{[O/H]}$ as a function of density for low and high mass galaxies. }
	\label{fig:sdss_mz}
\end{figure*}

Figure~\ref{fig:mz} shows that at small stellar masses, the MZ relation increases linearly with the logarithm of the stellar mass. At the low mass end, the metallicity is small and $Z(1-R) \ll Y_N$; Equation \ref{eq:dz} reduces to $dZ/dM_\ast \approx Y_N/M_g$. $Y_N$ is constant thus the metallicity is directly proportional to the stellar-to-gas mass ratio (this can also be seen by Taylor expanding Equation \ref{eq:ana}). At intermediate stellar masses the MZ relation begins to turnover and asymptotically approaches the upper metallicity limit at large stellar masses. This turnover occurs when the metallicity becomes large enough that an appreciable fraction of metals produced by massive stars become forever locked up in low mass stars and finally saturates when $Z(1-R) = Y_N$ (in Equation~\ref{eq:dz}, $dZ/dM_\ast = 0$). At this point, the amount of metals formed in high mass stars exactly equals the amount of metals forever locked up in low mass stars and the metallicity cannot increase further.  From Equation~\ref{eq:mz} and Equation~\ref{eq:m0mg} we see that the stellar mass where this turnover occurs, $M_0$, corresponds to the the stellar mass where the stellar-to-gas mass ratio is near unity.

\citetalias{zah14b} examine the redshift evolution of the MZ relation. The parameters $Z_0$ and $\gamma$ remain constant between $0.1 \lesssim z \lesssim 1.6$; $M_0$ increases with redshift. Thus, the metallicity plotted as a function of stellar mass normalized to $M_0$ is a universal, redshift-independent relation. Based on the UMR formulation, \citetalias{zah14b} conclude that this universal relation is a relation between metallicity and stellar-to-gas mass ratio. The evolution of $M_0$ simply reflects the redshift evolution of stellar-to-gas mass ratio. The larger $M_0$ at higher redshifts indicates that at a fixed stellar mass, galaxies contain more gas (Equation~\ref{eq:gas} and \ref{eq:m0}). Thus, the stellar masses and metallicities are proxies for the gas content of galaxies. 

We will use the UMR formulation to investigate the environmental dependence of the MZ relation. We fit the MZ relation for galaxies in different environments and infer the underlying physical properties from a comparison of the best-fit parameters with our chemical evolution model. As with the redshift evolution of the MZ relation, we will show that the variation of the MZ relation as a function of environment can be parameterized solely by a variation in $M_0$. We interpret this as a consequence of variation in the gas content of galaxies.

\section{The Environmental Dependence in the MZ-ENV Sample}
\label{sec:sdss}

Several studies of galaxies in the local universe suggest that the galaxies in higher density environments have higher metallicities \citep{mou07,coo08,ell09,pet12,pen14}. Here we examine this issue, quantifying the effect of the environment on the MZ relation using the UMR framework.

\subsection{The MZ Relation as a Function of Environment}
\label{sec:ori_env}

We sort the MZ-ENV sample into 5 equally populated bins of local density to investigate the environmental dependence of the MZ relation. Figure~\ref{fig:sdss_mz}a shows the MZ relation for the lowest and the highest density bins. Galaxies in high-density regions have higher metallicity than galaxies in the low-density regions.  However, the difference is small and is most prominent at the low mass end. This trend is similar to the mass dependent redshift evolution of the MZ relation \citep*[e.g.][]{zah11} and is a consequence of the fact that the saturation of metallicities occurs at larger stellar masses for galaxies residing in lower density environments. 

Figure~\ref{fig:sdss_mz}b shows the dependence of metallicity on environment for the MZ-ENV sample. For each galaxy, we measure the residual, $\Delta_{[O/H]}$, as the difference between the metallicity of the galaxy and the metallicity derived from the fit to the sample (see Section~\ref{sec:mz}). The median residual shows a weak correlation between metallicity and local density. We separate the MZ-ENV galaxies into two sub-samples segregated by stellar mass and plot the median $\Delta_{[O/H]}$ of each sub-sample. The environmental effect is more significant for low-mass galaxies. For high-mass galaxies, the $\Delta_{[O/H]}$ is nearly independent of environment. The overall behavior of the residuals agrees with previous studies \citep*{mou07,coo08}

\begin{figure*}
	\centering
	\includegraphics[width=0.9\textwidth]{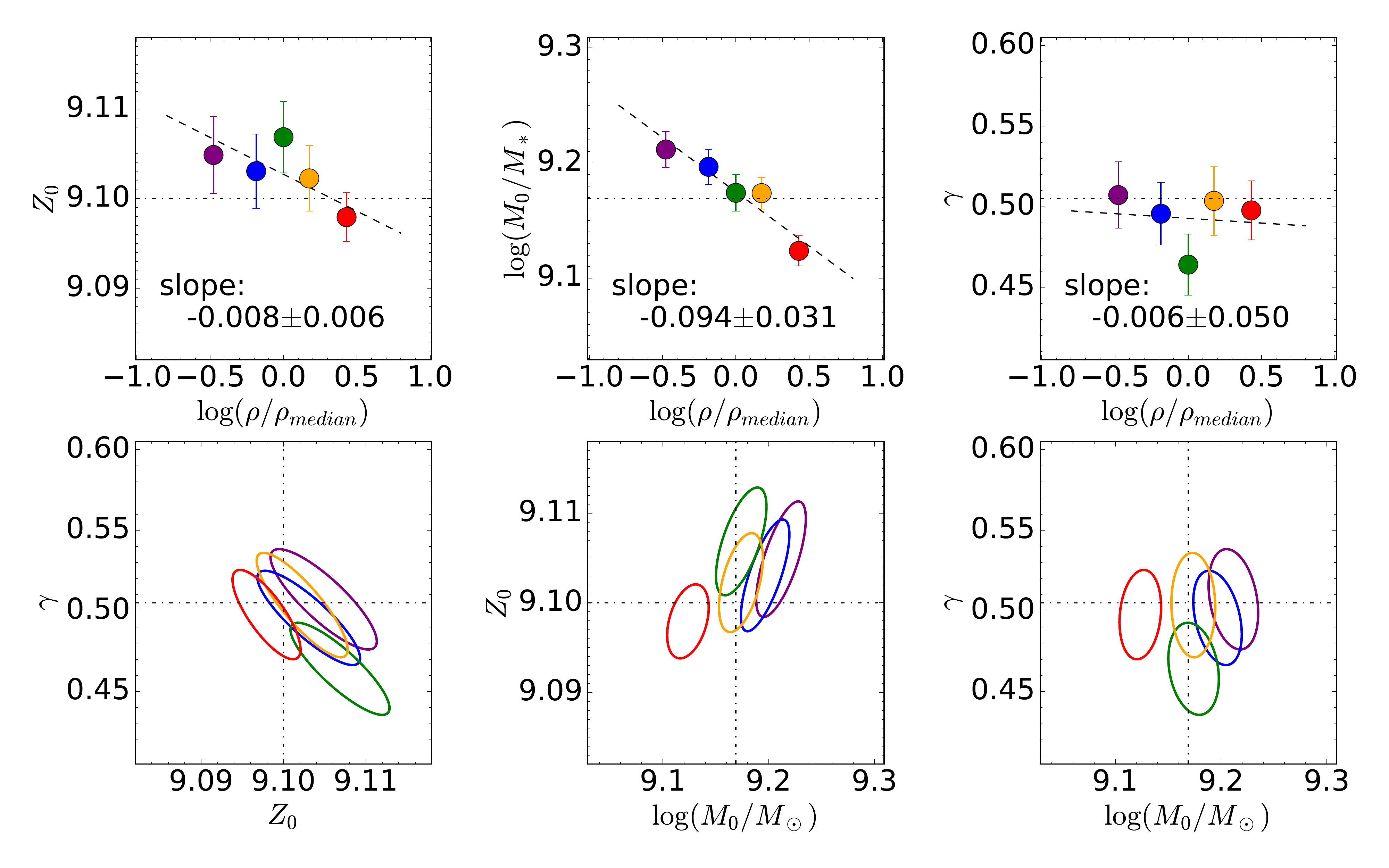}
	\caption{\textit{Upper:} Best-fit parameters in 5 density bins. The $M_0$ shows a clear evolution along with local density. On the contrary, the slopes of best-fit parameters as a function of local density are consistent with no environmental dependence for $Z_0$ and $\gamma$ ($\lesssim 1 \sigma$). \textit{Lower:} Covariances of three free parameters. The colour is the same as in the upper panels. The $M_0$ is largely independent from the other two parameters.}
	\label{fig:sdss_fit} 
\end{figure*}

We fit the MZ relation sorted in five bins of local density using our model (Equation~\ref{eq:mz}). Table~\ref{tab:sdss_mz_fit} and Figure~\ref{fig:sdss_fit} show the best-fit parameters, as well as the covariance among three parameters for the MZ relation fits. $Z_0$ and $\gamma$ for the five MZ relations are consistent ($\lesssim2\sigma$) with the best-fit value of the whole MZ-ENV sample. The density bin with $\log(\rho/\rho_{med})=0$ deviates from the mean values of $Z_0$ and $\gamma$ the most; $Z_0$ is higher and $\gamma$ is lower than the mean values by $\sim 2 \sigma$. We note that there is a strong covariance between $Z_0$ and $\gamma$, where lowering $Z_0$ and increasing $\gamma$ at the same time would also yield a reasonably good fit. Thus, the $>2\sigma$ offset in $Z_0$ and $\gamma$ seen for galaxies in the median density bin (green points) may in part be a consequence of the model covariance. On the contrary, $M_0$ has a significant dependence on local density. There is weak to no covariance between $M_0$ and $Z_0$, and $M_0$ and $\gamma$. The best-fit $M_0$ is not affected by varying $Z_0$ or $\gamma$. 

Figure~\ref{fig:sdss_fit} shows a linear fit to the best-fit model parameters as a function of local density. The slopes of the fit of $Z_0$ and $\gamma$ as a function local density are consistent with zero ($\lesssim 1\sigma$). The data suggest that $Z_0$ and $\gamma$ are independent of environment. The effect of the environment on the MZ relation can be quantified by the variation in $M_0$. In the UMR formulation, $M_0$ is directly related to the zero-point of the relation between gas mass and stellar mass (see Equation \ref{eq:m0}). This result implies that the environmental dependence on the MZ relation is mainly driven by variations in the zero-point of the relation between gas mass and stellar mass in star-forming galaxies. At fixed stellar mass, galaxies in high-density environments contain less gas, on average, as compared to galaxies in lower density environments.

\subsection{Quantifying the Environmental Effect on $M_0$ and the Gas Content}
\label{sec:m0rho}
\begin{figure}
\centering
\includegraphics[width=\columnwidth]{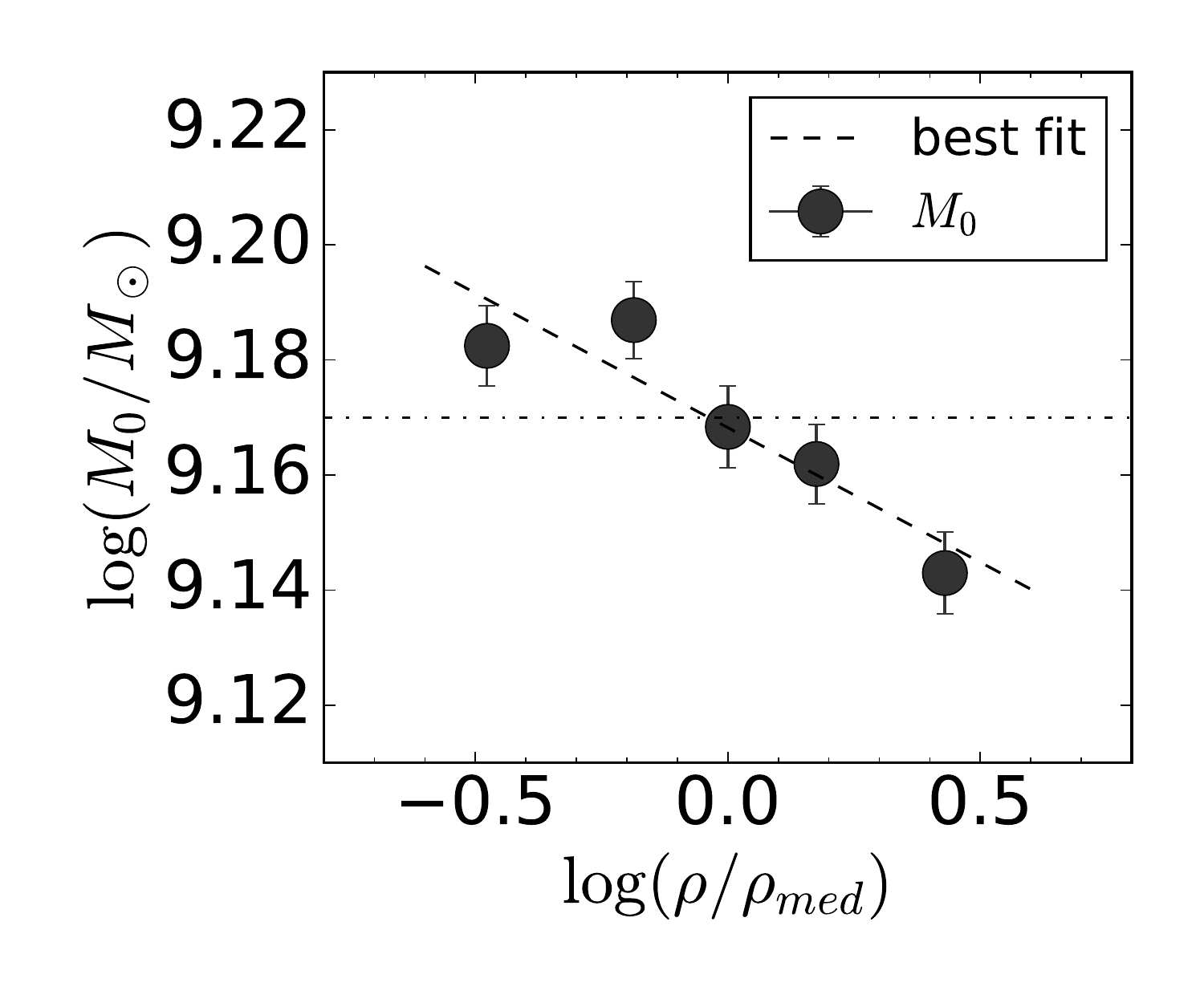}
\caption{Best-fit $M_0$ as a function of local densities, with $\gamma$ and $Z_0$ fixed. The dashed line is the linear best-fit. The $M_0$ decreases while the local density increases.}
\label{fig:m0fix_rho}
\end{figure}

In Section~\ref{sec:ori_env}, we show that $Z_0$ and $\gamma$ have no significant dependence on environment. To robustly quantify the environmental effect on $M_0$, we refit the MZ relation for each sub-sample but fix $Z_0$ and $\gamma$ to the best-fit value derived from the whole MZ-ENV sample. Thus, $M_0$ is the only free parameter. The best-fit $M_0$ is given in Table~\ref{tab:mz}. 
Figure~\ref{fig:m0fix_rho} shows $M_0$ as a function of density which is
\begin{equation}
M_0 = (-0.047 \pm 0.014) \times \log(\rho/\rho_{med}) + (9.168 \pm 0.006)
\label{eq:m0rho}
\end{equation}
 In the range of local density we probe [$-0.5 \lesssim \log(\rho/\rho_{med}) \lesssim 0.5$], $M_0$ varies by $\sim 0.05$~dex.

\begin{table}
	\caption{Best-fit model parameters in different environments with $Z_0$ and $\gamma$ fixed.}
	\label{tab:mz}
	\centering
	\begin{tabular}{cc}
		\hline
		\hline
		\multicolumn{2}{c}{$(Z_0,\gamma) = (9.100,0.505)$}  \\
		\hline 
		$\log(\rho/\rho_{med})$	& $\log (M_0/M_\odot)$ \\ 
		\hline
		Avg.   & $9.169\pm0.003$ \\
		\hline
		-0.48 & $9.182\pm0.007$  \\ 
		-0.19 & $9.187\pm0.007$  \\ 
		0.00 & $9.168\pm0.007$  \\ 
		0.18 & $9.162\pm0.007$  \\ 
		0.43 & $9.143\pm0.007$  \\ 
	
		\hline

	\end{tabular}
\end{table}

Using the UMR interpretation and Equations~\ref{eq:gas}, \ref{eq:gamma}, and \ref{eq:m0}, we can calculate the average gas content of star-forming galaxies as a function of stellar mass from the parameters we fit to the MZ relation:
\begin{equation}
M_g = M_0^\gamma M_\star^{1-\gamma}.
\label{eq:mg}
\end{equation}
The 0.05~dex variation in $M_0$ translates into $\sim 0.025$~dex ($\sim 6\%$) variation in gas mass. 

Measurements of the MZ relation suggest a subtle environmental dependence of the gas content in galaxies as a function of environment. We use Equation~\ref{eq:mg} to convert the best-fit $M_0$--$\rho$ relation shown in Figure~\ref{fig:m0fix_rho} and given by Equation \ref{eq:m0rho} to a $\Delta M_g$--$\rho$ relation. At a fixed stellar mass, $\Delta M_g$ is the gas mass at density $\rho$ relative to the average gas mass of a galaxy at the median density of the sample.
The relation is
\begin{equation}
\label{eq:mgfit_sdss}
\Delta_{Mg} = (-0.023 \pm 0.007) \times \log(\rho/\rho_{med}) - (0.001 \pm 0.003).
\end{equation}
This dependence of the ISM gas content of star forming galaxies on the environment accounts for the variation of the MZ relation we measure.

Previous studies have found that the SFR is proportional to the gas mass \citep{ken98,clu10,hua12}. Therefore, if our interpretation is correct, we anticipate a commensurate dependence of the SFR on the environment. At fixed stellar mass, star-forming galaxies in higher density environments should have lower SFRs because of their smaller gas reservoirs. 

We test the dependence of SFRs on environment by calculating
\begin{equation}
\Delta_{\Psi,i} = \log\left[ \Psi_i/  \overline{\Psi(M_{\ast,i})} \right].
\end{equation}
Here, $\Psi_i$ is the specific SFR (sSFR) of galaxy, $i$, and $\overline{\Psi(M_{\ast,i})}$ is the median sSFR rate for galaxies in the MZ-ENV sample at the same stellar mass. In practice, we calculate the $\overline{\Psi(M_{\ast,i})}$ by taking the median sSFR of galaxies in a narrow stellar mass range ($\Delta M_\ast \sim 0.05$~dex). If the ISM content of star-forming galaxies varies with environment as described in Equation~\ref{eq:mgfit_sdss}, we expect a smaller $\Delta_{\Psi,i}$ for galaxies in higher density environments.

\begin{figure}
	\centering
	\includegraphics[width=\columnwidth]{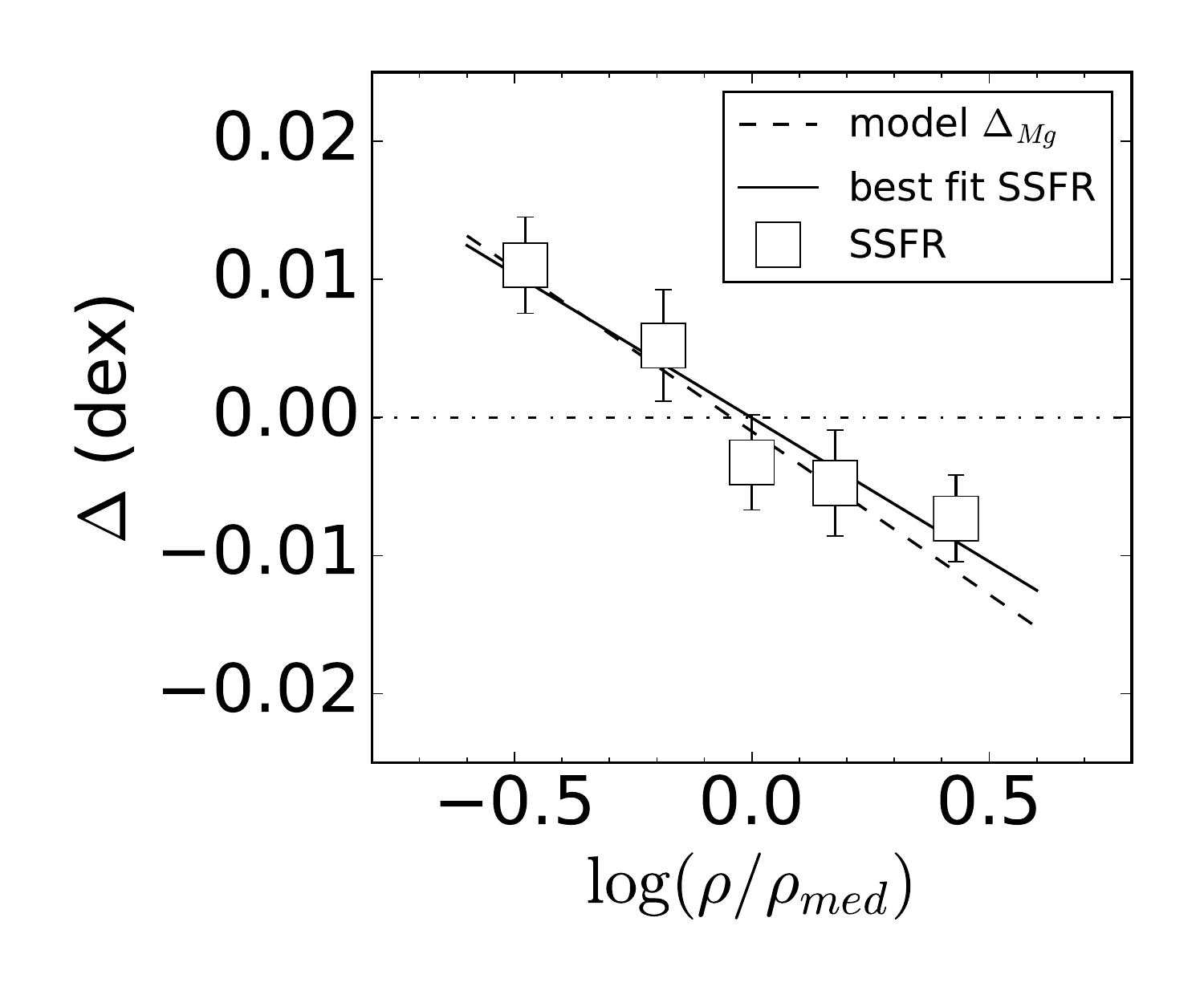}
	\caption{The residuals of the sSFR ($\Delta_\Psi$) as a function of local density. The residuals are calculated as the difference between the sSFR compared to the median value at given stellar mass (open squares). The solid line is a fit between the median $\Delta_\Psi$ and local density. The dashed line is the $\Delta M_g$ predicted from the variation in gas content inferred from the dependence of $M_0$ on environment assuming SFR$\propto M_g$ as a function of local density. The variation in gas mass inferred from the model agrees with the variation in sSFR.}
	\label{fig:dsfr_rho}
\end{figure}

Figure~\ref{fig:dsfr_rho} shows the median $\Delta_{\Psi,i}$ in bins of local density. The statistical uncertainty is determined by bootstrap resampling. The median $\Delta_{\Psi,i}$ in bins of local density is best-fit by
\begin{equation}
\label{eq:sfrfit_sdss}
\overline{\Delta_\Psi(\rho)} = (-0.021 \pm 0.006) \times \log(\rho/\rho_{med}) + (0.000 \pm 0.002)
\end{equation}
(solid line in Figure~\ref{fig:dsfr_rho}). The sSFR varies by $\sim 0.02$~dex as a function of local density. The dashed line in Figure~\ref{fig:dsfr_rho} is the $\overline{\Delta_\Psi}$ predicted from the variation in gas content inferred from measurements of the MZ relation (Equation \ref{eq:mgfit_sdss}) assuming that the SFR $\propto M_g$. The slope of the two relations (Equation~\ref{eq:sfrfit_sdss} and Equation~\ref{eq:mgfit_sdss}) are consistent. Thus, the variation in gas content as a function of local density we derive from our fit of the MZ relation is consistent with the variation of sSFR as a function of local density. Measurements of metallicity and sSFR are independent. The consistency supports our conclusion that the environmental dependence of the MZ relation is mainly a consequence of the gas content in star-forming galaxies. 

\section{Discussion}
\label{sec:dis}

Galaxies in higher local density regions have higher gas-phase metallicities on average. The effect is subtle and depends on stellar mass. The median metallicity of lower mass galaxies ($M_\star \lesssim 10^{10} M_\odot$) varies by $\sim 0.02$~dex over a local density range $-0.5 \lesssim \log(\rho/\rho_{med}) \lesssim 0.5$; the metallicities of massive galaxies ($M_\star \gtrsim 10^{10} M_\odot$) are unaffected by the large scale environment (Figure~\ref{fig:sdss_mz}). Our results agree with previous studies using similar data from the SDSS over the same range of environments \citep{mou07,coo08,ell09}. Within the UMR framework, this dependence of metallicity on environment is expected if gas content varies with environment. 

Early work investigating the impact of environment on gas content focused on galaxies residing in clusters. These studies demonstrate that cluster galaxies are gas deficient \citep[e.g.,][]{gio85,sol01,bot16}. Recent, blind H\,{\sc i} surveys of the galaxy population \citep[Arecibo Legacy Fast ALFA Survey, ALFALFA;][]{hay11} measure the gas content of galaxies over a broader range of environments. Galaxies in denser environments, in general, have less gas \citep{fab12,cat13,ode16}. Thus, the gas content of galaxies as probed by direct measurements appears to be anti-correlated with local density. 

Metallicity is also reported to be anti-correlated with both the SFR and gas-content of galaxies. At fixed stellar mass, low metallicity galaxies on average have more H\,{\sc i} gas and have higher star formation rates \citep{bot13,man10,man11}. From these observations, a galaxy at low density region is expected to have more gas and, as a result, lower metallicity. The UMR framework provides a theoretical framework for interpreting the relation between stellar mass, metallicity and environment in terms of gas-content.

Metallicity measured from strong emission lines of H\,{\sc ii} regions is essentially a SFR-weighted average and is only sensitive to gas in the ISM \citep{bot16}. Thus, the use of metallicities as a proxy for gas content is complementary to direct measurements of cold gas which may be sensitive to gas distributed more broadly than the ISM of galaxies. H\,{\sc i} deficient galaxies in clusters have higher metallicity \citep{ski96,dor06,zha09,ell09}. These studies demonstrate that in clusters the deficiency of HI directly impacts the gas-content of the interstellar medium. We extend these results based on analysis of cluster galaxies to a larger range of environments; gas-content is weakly correlated with the large scale environment and this explains the subtle dependence of metallicities and specific star formation rates of galaxies on environment.

Currently, direct measurements of H\,{\sc i} gas mass are still largely confined to the local universe due to the sensitivity of observing facilities. For example, the ALFALFA survey provides more than ten thousand sources, but it only reaches out to $z\sim 0.05$. Measurements of metallicity provide an alternative to probe the gas content in star-forming galaxies. Because this method measures strong optical emission lines, it can be applied to a large number of galaxies at higher redshifts with reasonable observing resource. With future large-scale spectroscopic surveys such as the Subaru Prime Focus Spectrograph (PFS) survey \citep{tak14}, this method can potentially estimate the gas content in star-forming galaxies out to higher redshifts. 
 
For the UMR model, the saturation metallicity $Z_0$ depends on the nucleosynthetic yield and the net fraction of the oxygen formed in massive stars that is expelled from galaxies in outflows; $\gamma$ is set by the slope of the stellar mass --- gas mass relation. \citetalias{zah14b} show that $Z_0$ and $\gamma$ do not depend on redshift. Here we show that these parameters do not depend on environment either. The insensitivity of these parameters to redshift and environment provides constraints for models of galaxy formation and evolution which include star formation, feedback and gas flows. In particular, the slope of the relation between stellar mass and gas mass and the fraction of oxygen formed by galaxies which is expelled from the interstellar medium appear to be independent of redshift and the large scale environment.

\section{Summary}
\label{sec:sum}
We examine the MZ relation as a function of environment based on analysis of $\sim40,000$ galaxies in the SDSS. We show that the gas-phase metallicity has a weak dependence on the environment. 

The MZ relation is a power-law at low stellar masses and saturates at high stellar masses. The shape of the MZ relation does not depend on environment. The environmental dependence is completely parameterized by variation of the turnover stellar mass where the MZ relation saturates; the turnover mass is smaller in high density environments. At stellar masses below the turnover stellar mass, galaxies in higher local density regions have on average higher metallicity. Above the turnover mass the metallicity saturates and thus is independent of environment. 

We interpret the environmental dependence of the MZ relatoin using the analytical chemical evolution model of \citetalias{zah14b}. Based on comparing the best-fit parameters with the model we show that the turnover of the MZ relation depends on the gas-content of galaxies. Thus the gas content in star-forming galaxies is the most important parameter for understanding the impact of environment on the metallicity. The specific star formation has a corresponding dependence on the environment supporting our interpretation. We show that a $\sim 5\%$ change in gas content over the density regime probed in this study can explain the measured dependence of the metallicity and specific star formation rate on the environment. 

Modeling the MZ relation provides a means to probe the average gas content in star-forming galaxies. This method can be applied to galaxies at higher redshifts. With large spectroscopic surveys, this method may be sensitive to few percent difference in gas mass, thus providing a powerful tool to study the gas content in star-forming galaxies through cosmic time. 

\section*{Acknowledgments}

We thank the referee for carefully reading the manuscript and providing constructive comments. HJZ is supported by the Clay Fellowship and MJG is supported by the Smithsonian Institution. This research was supported by the Munich Institute for Astro- and Particle Physics (MIAPP) of the DFG cluster of excellence "Origin and Structure of the Universe".

Funding for SDSS-III has been provided by the Alfred P. Sloan Foundation, the Participating Institutions, the National Science Foundation, and the U.S. Department of Energy Office of Science. The SDSS-III web site is \url{http://www.sdss3.org/}.

SDSS-III is managed by the Astrophysical Research Consortium for the Participating Institutions of the SDSS-III Collaboration including the University of Arizona, the Brazilian Participation Group, Brookhaven National Laboratory, Carnegie Mellon University, University of Florida, the French Participation Group, the German Participation Group, Harvard University, the Instituto de Astrofisica de Canarias, the Michigan State/Notre Dame/JINA Participation Group, Johns Hopkins University, Lawrence Berkeley National Laboratory, Max Planck Institute for Astrophysics, Max Planck Institute for Extraterrestrial Physics, New Mexico State University, New York University, Ohio State University, Pennsylvania State University, University of Portsmouth, Princeton University, the Spanish Participation Group, University of Tokyo, University of Utah, Vanderbilt University, University of Virginia, University of Washington, and Yale University.

\appendix
\section{Testing Metallicity Diagnostics and Sample selection}
\label{app}

We test how different metallicity diagnostics and sample selections affect our result. In this Appendix, we show the results with
\begin{itemize}
	\item O3N2 and N2 diagnostics;
	\item galaxies in different redshift ranges;
	\item more stringent S/N criteria;
	\item more stringent fiber covering fraction requirements.
\end{itemize} 
The results of these various analysis are consistent with the results presented in the main text. In particular, we conclude that only $M_0$ varies as function of overdensity. The measured dependence of $M_0$ on overdensity is consistent with results presented in the main text; the errors are larger due to smaller sample size. 

\subsection{Metallicity diagnostics}
\citet{pet04} used electron temperature-based metallicities from H\,{\sc ii} regions to emperically calibrate two metallicity diagnostics, the N2 and O3N2 indices, where $N2 \equiv \log(\mbox{[N\,{\sc ii}]}\lambda6583/\mbox{H}\alpha)$, and $O3N2 \equiv \log\{(\mbox{[O\,{\sc iii}]}\lambda5007/\mbox{H}\beta)/(\mbox{[N\,{\sc ii}]}\lambda6583/\mbox{H}\alpha ) \}$. 
 
\begin{figure*}
	\includegraphics[width=\textwidth]{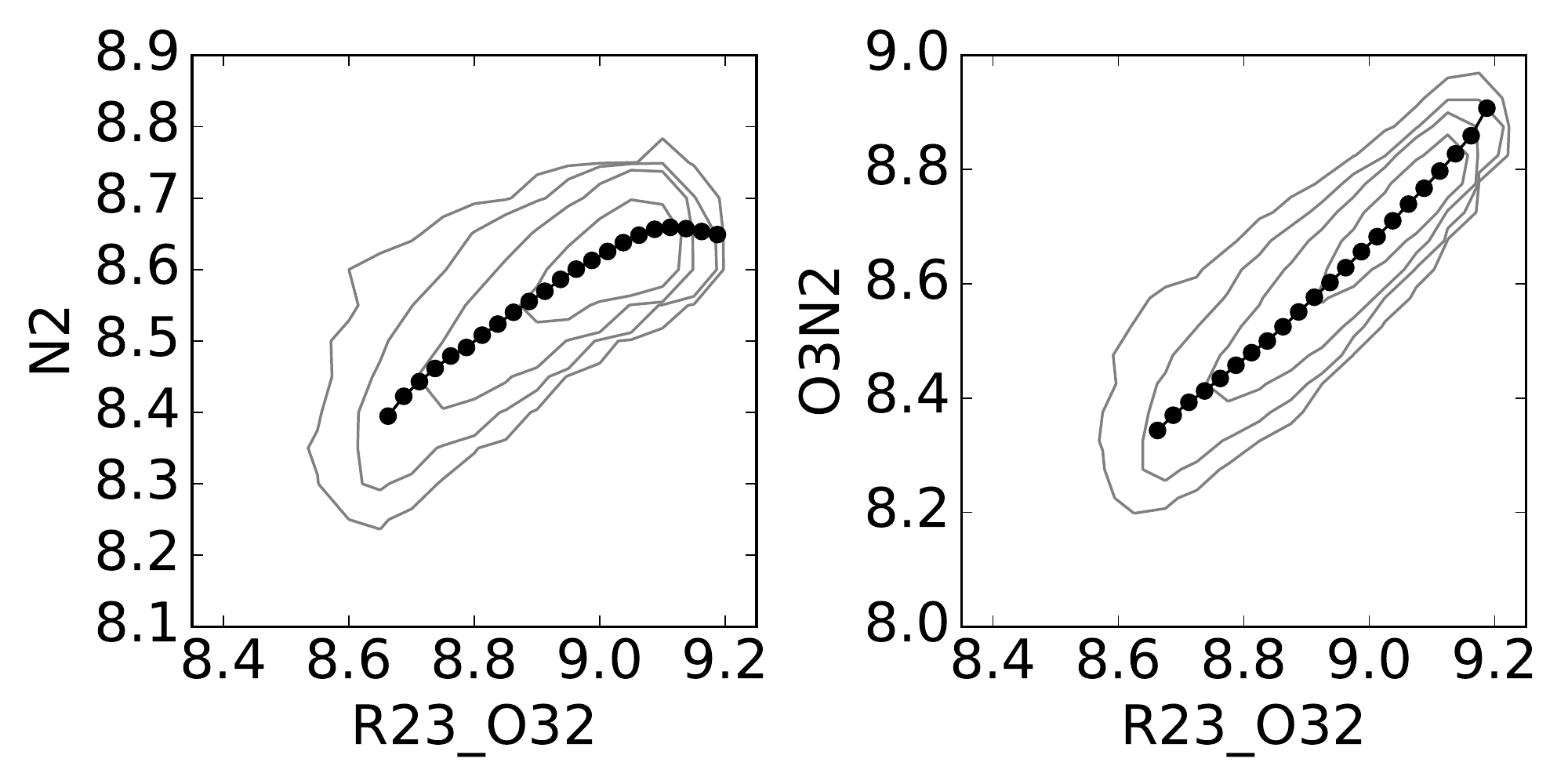}
	\caption{Comparison between metallicities derived from different diagnostics. Contours represent for galaxy density on the 0.05 by 0.05 metallicity grids. Large black dots are the median value of N2 or O3N2 metallicity at fixed R23 metallicity. The N2 metallicity artificially saturates at high metallicities. }
	\label{fig:N2}
\end{figure*}

\begin{figure*}
	\includegraphics[width=\textwidth]{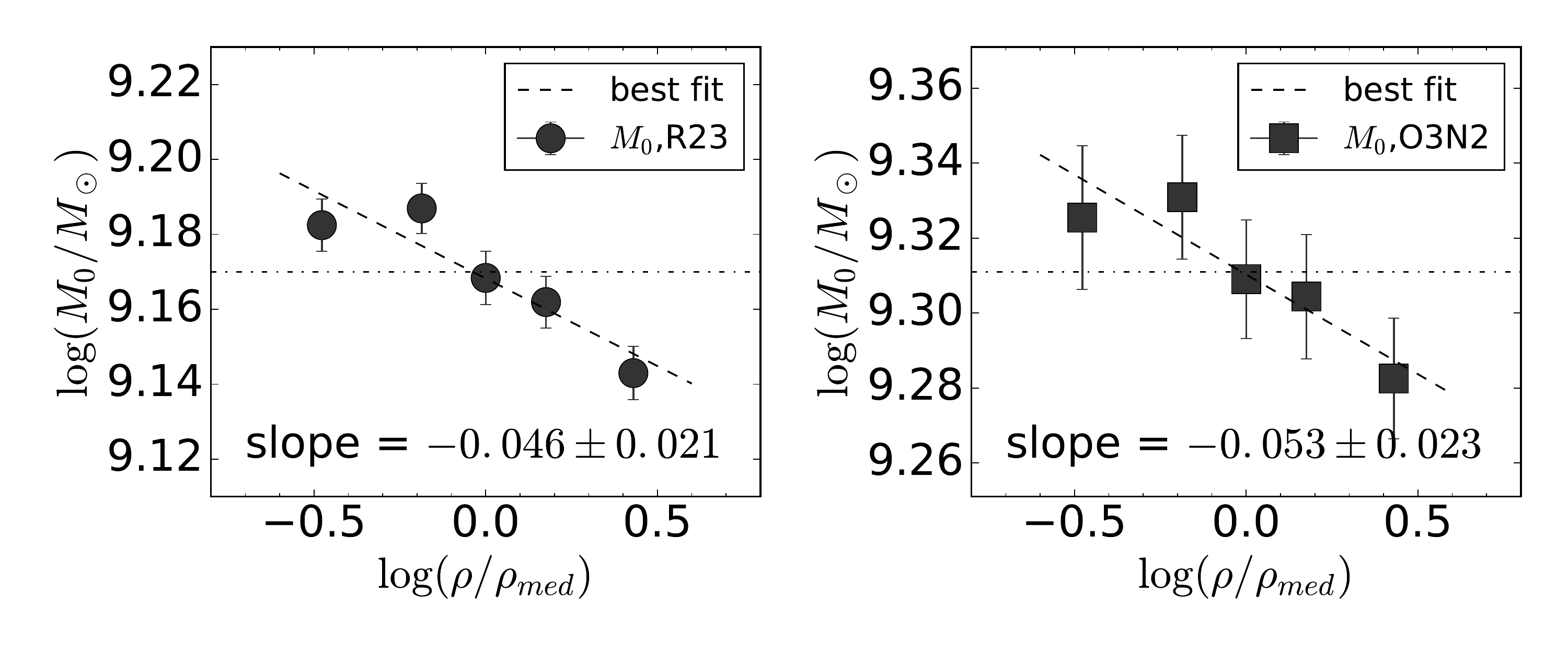}
	\caption{Comparing the best-fit $M_0$ as a function of environment using R23 and O3N2 diagnostics, with $Z_0$ and $\gamma$ fixed to the best-fit value using the whole sample. The left panel is the same as Figure~\ref{fig:m0fix_rho}.}
	\label{fig:O3N2_M0}
\end{figure*}

Above solar metallicity the N2 index saturates and is not sensitive to metallicity \citep{pet04}. Fig.~\ref{fig:N2} compares the N2 metallicity of our sample galaxies to the R23 diagnostics. The flattening at high metallicities demonstrates the artificial saturation of N2 metallicity. Saturation of the N2 diagnostic masks the turnover in the MZ relation and thus can not be used to measure the turnover mass in the MZ relation. On the other hand, the O3N2 calibration is sensitive at high metallicity end (right panel of Fig.~\ref{fig:N2}). We thus test our result with the O3N2 calibration. Fig.~\ref{fig:O3N2_M0} compares the best-fit $M_0$ using R23 and O3N2 methods. The two methods give consistent environmental dependence for $M_0$.

\subsection{Redshift}

\begin{figure*}
	\includegraphics[width=\textwidth]{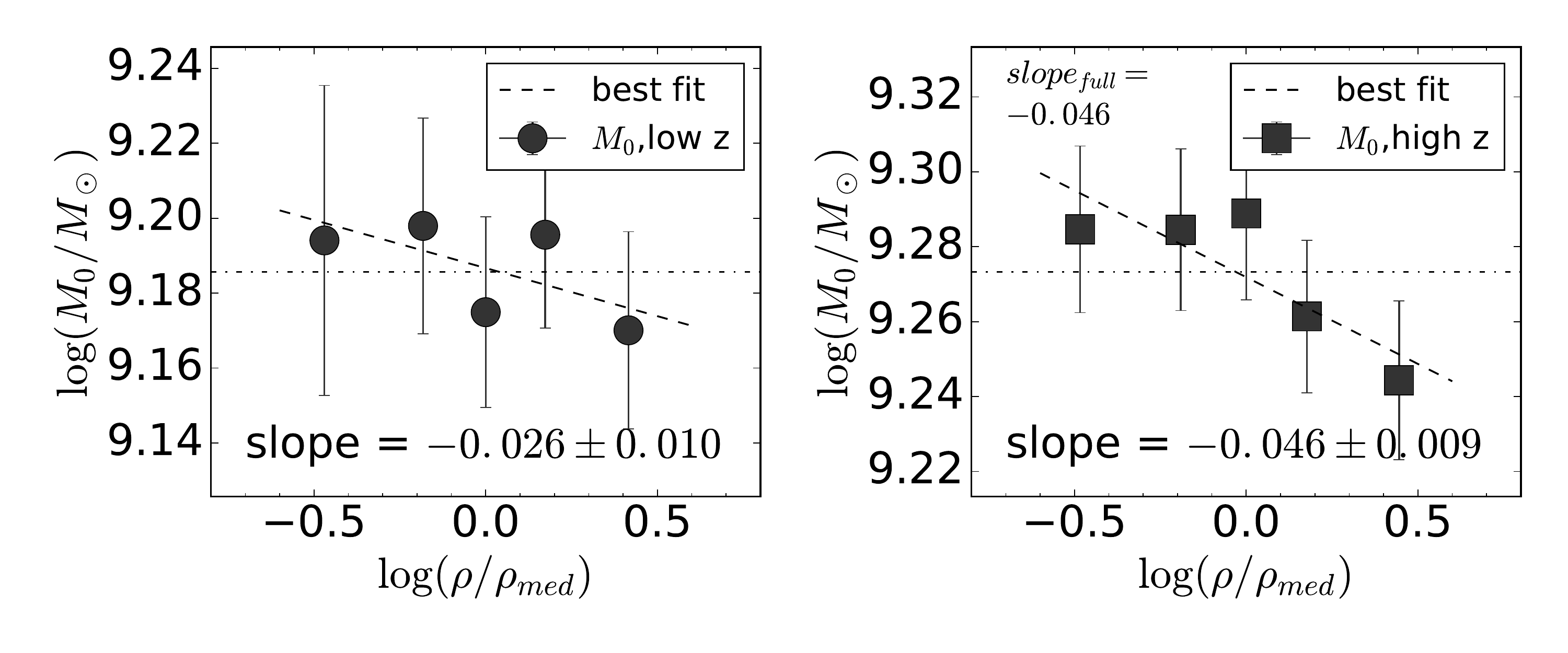}
	\caption{Best-fit $M_0$ for the low-z subsample ($z\leq0.0765$, left panel) and the high-z subsample ($z>0.0765$, right panel), respectively. The $Z_0$ and $\gamma$ are fixed to the best-fit value. The slope of the best-fit $M_0-\rho$ relation from the MZ-env sample is labeled at the corner of the right panel.  }
	\label{fig:z_param}
\end{figure*}

We separate the MZ-env sample into two equal-size redshift bins. The result is plotted in Fig.~\ref{fig:z_param}. The large uncertainty in the low-z subsample is a result of lacking high mass galaxies. In the low-z sample, only $< 10\%$ of galaxies have $\log(M_\odot/M_\star) > 10$. Therefore, the shape of the MZ relation is poorly constrained but consistent with results presented in Section~\ref{sec:m0rho}.

\subsection{S/N ratio}

\begin{figure*}
	\includegraphics[width=\textwidth]{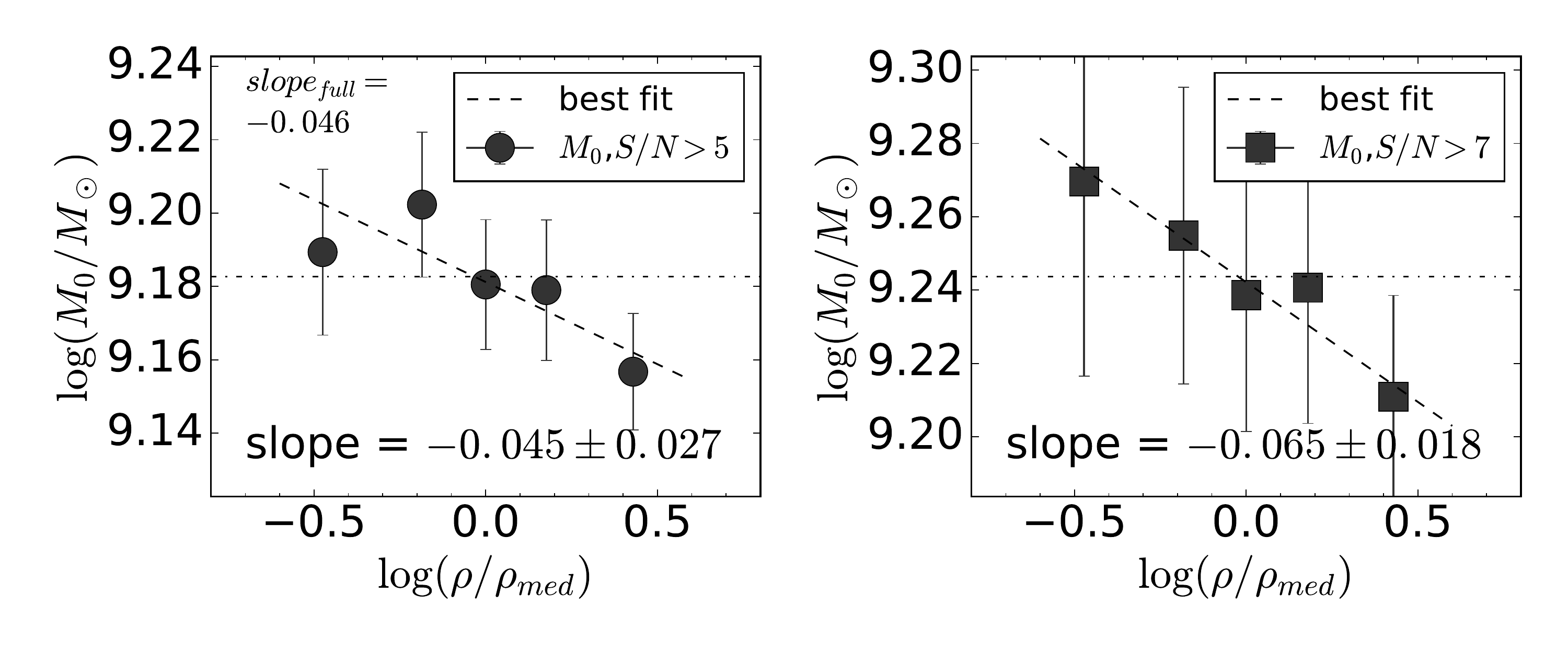}
	\caption{Best-fit parameters for $S/N>5$ (left panel) and $S/N>7$ (right panel). The sample size is $\sim 90\%$ and $\sim 70\%$ of the MZ-env sample for $S/N>5$ and $S/N>7$, respectively. The $Z_0$ and $\gamma$ are fixed to the best-fit value. The slope of the best-fit $M_0-\rho$ relation from the MZ-env sample is labeled at the corner of the left panel.}
	\label{fig:sn_param}
\end{figure*}

We apply an $S/N>3$ cut for the MZ-env sample. Here we show two results with $S/N>5$ and $S/N>7$ (Fig.~\ref{fig:sn_param}). 
We note that applying an S/N cut is similar to applying a mass-dependent cut on sSFR. Because of the strong dependence of metallicity on sSFR, a stringent S/N cut potentially distorts the shape of the MZ relation \citep[see][]{fos12}. The large uncertainty with $S/N>7$ cut is mainly due to lack of high-mass galaxies in the sample. 

\subsection{Covering fraction}

\begin{figure*}
	\includegraphics[width=\textwidth]{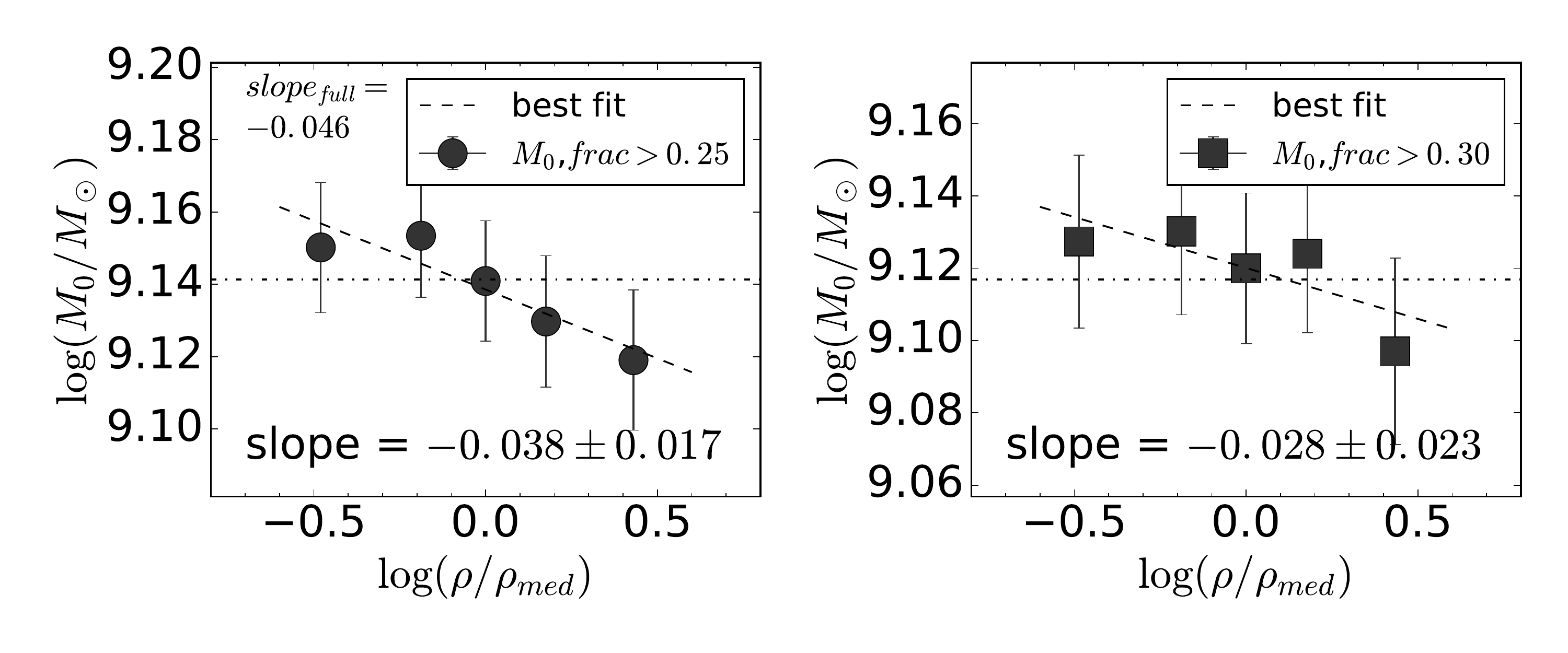}
	\caption{Best-fit parameters for different fiber covering fractions. The sample size is $\sim 75\%$ and $\sim 50\%$ of the MZ-env sample for covering fraction $>25\%$ (left panel) and $>30\%$ (right panel), respectively. The $Z_0$ and $\gamma$ are fixed to the best-fit value. The slope of the best-fit $M_0-\rho$ relation from the MZ-env sample is labeled at the corner of the left panel.}
	\label{fig:f_param}
\end{figure*}

Imposing a minimum covering fraction potentially introduces a selection bias on galaxy structure. The 20\% minimum covering fraction is to ensure the metallicity measured from fiber is representative of the globle value \citep{kew05}. Fig.~\ref{fig:f_param} shows the result with minimum fiber covering fraction of 25\% and 30\%.

\end{CJK*}

\begin{thebibliography}{}
	\bibitem[Abazajian et al.(2009)]{aba09} Abazajian, K. N., et al., 2009, ApJS, 182, 543
	\bibitem[Aihara et al.(2011)]{aih11} Aihara, H., et al. 2011, ApJS, 193, 29
	\bibitem[Alam et al.(2015)]{ala15} Alam, S., Albareti, F. D., Allende Prieto, C. 2015, APJS, 219, 12
	\bibitem[Alloin et al.(1979)]{all79} Alloin, D., Collin-Souffrin, S., Joly, M., Vigroux, L., 1979, A\&A, 78, 200
	\bibitem[Arnouts et al.(1999)]{arn99} Arnouts, S., Cristiani, S., Moscardini, L., Matarrese, S., Lucchin, F., Fontana, A., Giallongo, E., 1999, MNRAS, 310, 540
	\bibitem[\protect\citeauthoryear{Baldwin, Phillips, \& Terlevich}{Baldwin et al.}{1981}]{bal81} Baldwin, J. A., Phillips, M. M., \& Terlevich, R., 1981, PASP, 93, 5
	\bibitem[Bekki et al.(2001)]{bek01} Bekki, K., Couch, W. J., Drinkwater, M. J., Gregg, M. D., 2001, ApJL, 557, 39
	\bibitem[Bigiel \& Blitz(2012)]{bie12} Bigiel, F. \& Blitz, L., 2012, ApJ, 756, 183
	\bibitem[Bigiel et al.(2010)]{bie10} Biegiel, F., Leroy, A., Walter, F., Blitz, L., Brinks, E., de Blok, W. J. G., Madore, B., 2010, AJ, 140, 1194
	\bibitem[Blanton \& Roweis(2007)]{bla07} Blanton, M. R., Roweis, S., 2007, AJ, 133, 734 
	\bibitem[Blanton et al.(2003)]{bla03} Blanton, M., et al., 2003, ApJ, 594, 186
	\bibitem[Borthakur et al.(2015)]{bor15} Borthakur, S., et al., 2015, ApJ, 813, 46
	\bibitem[Bosma(1978)]{bos78} Bosma, A., 1978, PhD Thesis, Groningen Univ.
	\bibitem[Boselli \& Gavazzi(2006)] Boselli, A., Gavazzi, G., 2006, PASP, 118, 517
	\bibitem[Bothwell et al.(2016)]{bot16} Bothwell, M. S., Maiolino, R., Peng, Y., Cicone, C., Griffith, H., Wagg, J., 2016, MNRAS, 455, 1156
	\bibitem[Bothwell et al.(2013)]{bot13} Bothwell, M. S., Maiolino, R., Kennicutt, R., Cresci, G., Mannucci, F., Marconi, A., Cicone, C., 2013, MNRAS, 433, 1425
	\bibitem[Bravo-Alfaro et al.(2000)]{bra00} Bravo-Alfaro, H., Cayatte, V., van Gorkom, J. H., Balkowski, C., 2000, AJ, 119, 580
	\bibitem[Brinchmann et al.(2004)]{bri04} Brinchmann, J., Charlot, S., White, S. D. M., Trmonti, C., Kauffmann, G., Heckman, T., Brinkmann J. 2004, MNRAS, 351, 1151
	\bibitem[Bruzual \& Charlot(2003)]{bc03} Bruzual, G. \& Charlot, S. 2003, MNRAS, 344, 1000
	\bibitem[Calzetti et al.(2000)]{cal00} Calzetti, D., et al., 2000, ApJ, 533, 682
	\bibitem[\protect\citeauthoryear{Cardelli, Clayton, \& Mathis}{Cardelli et al.}{1989}]{car89} Cardelli, J. A., Clayton, G. c., Mathis, J. S. 1989, ApJ, 345, 245
	\bibitem[Catinella et al.(2013)]{cat13} Catinella, B., et al. 2013, MNRAS, 436, 34
	\bibitem[Chabrier(2003)]{cha03} Chabrier, G. 2003, PASP, 115, 763
	\bibitem[Charlot \& Longhetti(2001)]{cha01} Charlot, S., Longhetti, M. 2001, MNRAS, 323, 887
	\bibitem[\protect\citeauthoryear{Choi, Park, \& Vogeley}{Choi et al.}{2007}]{cho07} Choi, Y.-Y., Park, C., Vogeley, M. S., 2007, ApJ, 658, 884
	\bibitem[Chung et al.(2009)]{chu09} Chung, A., van Gorkom, J. H., Kenney, J. D. P., Crowl, H., Vollmer, B., 2009, AJ, 138, 1741 
	\bibitem[Cluver et al.(2010)]{clu10} Cluver, M. E., Jarrett, T. H., Krann-Korteweg, R. C., Koribalski, B. S., Appleton, P. N., Melbourne, J., Emonts, B., Woudt, P. A., 2010, ApJ, 725, 1550
	\bibitem[Cooper et al.(2008)]{coo08} Cooper, M. C., Tremonti, C. A., Newman, J. A., Zabludoff, A. I., 2008, MNRAS, 390, 245
	\bibitem[\protect\citeauthoryear{de Lapparent, Geller, \& Huchra}{de Lapparent et al.}{1988}]{del88} de Lapparent, V., Geller, M. J, Huchra, J. P., 1988, ApJ, 332, 44
	\bibitem[Dors \& Copetti(2006)]{dor06} Dors, O. L., Copetti, M. V. F., 2006, A\&A, 452, 473
	\bibitem[Diaferio(1999)]{dia99} Diaferio, A., 1999, MNRAS, 309, 610
	\bibitem[Dressler(1980)]{dre80} Dressler, A. J., 1980, ApJ, 236, 351
	\bibitem[Ellison et al.(2009)]{ell09} Ellison, S. L., Simard, L., Cowan, N. B., Baldry, I. K., Patton, D. R., McConnachie, A. W., 2009, MNRAS, 396, 1257
	\bibitem[Fabello et al.(2012)]{fab12} Fabello, S., Kauffmann, G., Catinella, B., Li, C., Giovanelli, R., Haynes, M. P. 2012, MNRAS, 427, 2841
	\bibitem[Fabricant et al.(2005)]{fab05} Fabricant, D., et al. 2005, PASP, 117, 1411
	\bibitem[Fabricant et al.(2008)]{fab08} Fabricant, D., Kurtz, M. J., Geller, M. J., Caldwell, N., Woods, D., Dell'Antonio, L., PASP, 120, 1222
    \bibitem[Foster et al.(2012)]{fos12} Foster, C., et al., 2012, A\&A, 547, A79
    \bibitem[Fujita \& Goto(2004)]{fuj04} Fujita, Y., Goto, T., 2004, PASJ, 56, 621
	\bibitem[Geller et al.(2010)]{gel10} Geller, M., J., Kurtz, M. J., Dell'Antonio, I. P., Ramella, M., Fabricant, D. G., 2010, ApJ, 709, 832
	\bibitem[Giovanelli \& Haynes(1985)]{gio85} Giovanelli, R., Haynes, M. P., 1985, ApJ, 292, 404 
	\bibitem[Gunn \& Gott(1972)]{gun72} Gunn J. E., Gott, J. R., ApJ, 176, 1
	\bibitem[Hashimoto et al.(1998)]{has98} Hashimoto, Y., Oemler, Jr., A., Lin, H., Tucker, D. L. 1998, ApJ, 499, 589
	\bibitem[Haynes et al.(2011)]{hay11} Haynes, M. P., et al. 2011, AJ, 142, 170
	\bibitem[Hogg et al.(2004)]{hog04} Hogg, D. W., et al. 2004, ApJ, 601, L29
	\bibitem[Huang et al.(2012)]{hua12} Huang, S., Haynes, M. P., Giovanelli, R., Brinchmann, J., Stierwalt, S., Neff, S. G., 2012, AJ, 143, 133
	\bibitem[Hwang et al.(2010)]{hwa10} Hwang, H. S., Elbaz, D., Lee, J. C., et al. 2010, A\&A, 522, 33
	\bibitem[Ilbert et al.(2005)]{ilb05} Ilbert, O., et al., 2005, A\&A, 439, 863
	\bibitem[Ilbert et al.(2006)]{ilb06} Ilbert, O., et al., 2006, A\&A, 457, 841
	\bibitem[Iovino et al.(2010)]{iov10} Iovino, A., et al., 2010, A\&A, 509, 40
	\bibitem[Jaff\'{e} et al.(2012)]{jaf12} Jaff\'{e}, Y. L., Poggianti, B. M., Verheijen, M. A. W., Deshev, B. Z., van Gorkom, J. H., 2012, ApJL, 756, 28
	\bibitem[Kauffmann et al.(2004)]{kau04} Kauffmann, G., White, S. D. M., Heckman, T. M., M\'{e}nard, B., Brinchmann, J., Charlot, S., Tremonti, C., Brinkmann, J. 2004,
	MNRAS, 353, 713
	\bibitem[Kawata \& Mulchaey(2008)]{kaw08} Kawata, D., Mulchaey, J. S., 2008, ApJL, 672, 103
	\bibitem[Kennicut(1998)]{ken98} Kennicut, R. C., Jr. 1998, ARA\&A, 36, 189
	\bibitem[Kewley \& Ellison(2008)]{kew08} Kewley, L. J., Ellison, S. L., 2008, ApJ, 681, 1183
	\bibitem[\protect\citeauthoryear{Kewley, Jansen, \& Geller}{Kewley et al.}{2005}]{kew05} Kewley, L. J., Jansen, R. A., Geller, M. J., 2005, PASJ, 117, 227
	\bibitem[Kewley et al.(2006)]{kew06} Kewley, L. J., Groves, B., Kauffmann, G., Heckman, T. 2006, MNRAS, 372, 961
	\bibitem[Kewley et al.(2010)]{kew10} Kewley, L. J., Rupke, D., Zahid, H. J., Geller, M. J., Barton, E. J. 2010, ApJL, 721, 48
	\bibitem[Kobayashi et al.(2006)]{kob06} Kobayashi, C., Umeda, H., Nomoto, J., Tominaga, N., Ohkubo, T. 2006, ApJ, 653, 1145
	\bibitem[Kobulnicky \& Kewley(2004)]{kk04} Kobulnicky, H. A., Kewley, L. J. 2004, ApJ, 617, 240
	\bibitem[Kulas et al.(2013)]{kul13} Kulas, K. R., et al. 2013, ApJ, 774, 130
	\bibitem[Larson(1972)]{lar72} Larson, R. B. 1972, NPhS, 236, 7
	\bibitem[\protect\citeauthoryear{Larson, Tinsley, \& Caldwell}{Larson et al.}{1980}]{lar80} Larson, R. B., Tinsley, B. M., Caldwell, C. N., 1980, ApJ, 237, 692
	\bibitem[Lequeux et al.(1979)]{leq79} Lequeux, J., Peimbert, M., Rayo, J. F., Serrano, A., Torres-Peimbert, S. 1979, A\&A, 80, 155
	\bibitem[Leroy et al.(2008)]{ler08} Leroy, A. K., Walter, F., Brinks, E., Bigiel, F., de Blok, W. J. G., Madore, B., \& Thornley, M. D., 2008, AJ, 136, 2782
	\bibitem[Lilly et al.(2013)]{lil13} Lilly, S. J., Carollo, C. M., Pipino, A., Renzini, A., Peng, Y. 2013, ApJ, 772, 119
	\bibitem[\protect\citeauthoryear{Liivam\"{a}gi, Tempel, \& Saar}{Liivam\"{m}agi et al.}{2012}]{lii12} Liivam\"{a}gi, L. J., Tempel, E., Saar, E. 2012, A\&A, 539, A80
	\bibitem[Mannucci et al.(2010)]{man10} Mannucci, F., Cresci, G., Maiolino, R., Marconi, A., Gnerucci, A., 2010, MNRAS, 408, 2115 
	\bibitem[\protect\citeauthoryear{Mannucci et al.}{Mannucci, Salvaterra, \& Campisi}{2011}]{man11} Mannucci, F., Salvaterra, R., Campisi, M. A., 2011, MNRAS, 414, 1263 
	\bibitem[Markwardt(2009)]{mar09} Markwardt, C. B. 2009, in ASP Conf. Ser. 411, Astronomical Data Analysis
	Software and Systems XVIII, ed. D. A. Bohlender, D. Durand, \& P. Dowler
	(San Francisco, CA: ASP), 251
	\bibitem[McCarthy et al.(2008)]{mcc08} McCarthy, I. G., et al., 2008, MNRAS, 383, 593
	\bibitem[McGee et al.(2009)]{mcg09} McGee, S. L., Balogh, M. L., Bower, R. G., Font, A. S., McCarthy, I. G., 2009, MNRAS, 400, 937
	\bibitem[Mouhcine et al.(2007)]{mou07} Mouhcine, M., Baldry, I. K., Bamford, S. P., 2007, MNRAS, 382, 801
	\bibitem[Oemler(1974)]{oem74} Oemler, A. 1974, ApJ, 194, 1
	\bibitem[Odekon et al.(2016)]{ode16} Odekon, M. C., et al. 2016, ApJ, 824, 110
	\bibitem[Oppenheimer \& Dav\'{e}(2006)]{opp06} Oppenheimer, B. D., Dav\'{e}, R., 2006, MNRAS, 373, 1265
	\bibitem[Oppenheimer \& Dav\'{e}(2008)]{opp08} Oppenheimer, B. D., Dav\'{e}, R., 2006, MNRAS, 387, 577
	\bibitem[Osterbrock (1989)]{ost89} Osterbrock, D. E. (ed.) 1989, Astrophysics of Gaseous Nebulae and Active 	Galactic Nuclei (Mill Valley, CA: University Science Books)
	\bibitem[Padmanabhan et al.(2008)]{pad08} Padmanabhan, N. et al. 2008, ApJ, 674, 1217
	\bibitem[Pagel \& Patchett(1975)]{pag75} Pagel, B. E. J., \& Patchett, B. E. 1975, MNRAS, 172, 13
	\bibitem[Papatergis et al.(2012)]{pap12} Papastergis, E., Cattaneo, A., Huang, S., Giovanelli, R., Haynes, M. P. 2012, ApJ, 759, 138
	\bibitem[Peeples \& Shankar(2011)]{pee11} Peeples, M. S., Shankar, F. 2011, MNRAS, 417, 2962
	\bibitem[Peeples et al.(2014)]{pee14} Peeples, M. S., Werk, J. K., Tumlinson, J., Oppenheimer, B. D., Prochaska, J., X., Katz, N., Weinberg, D. H. 2014, ApJ, 786, 54
	\bibitem[\protect\citeauthoryear{Petropoulou et al.}{Petropoulou, V\'{i}lchez, \& Iglesias-P\'{a}ramo}{2012}]{pet12} Petropoulou, V., V\'{i}lchez, J., Iglesias-P\'{a}ramo, J. 2012, ApJ, 749, 133
	\bibitem[Pettini \& Pagel(2004)]{pet04} Pettini, M., Pagel, B. E. J., 2004, MNRAS, 348, 59
	\bibitem[Peng \& Maiolino(2014)]{pen14} Peng, Y.-J., Maiolino, R., 2014, MNRAS, 438, 262
	\bibitem[Rasmussen et al.(2012)]{ras12} Rasmussen, J., Mulchaey, J. S., Bai, Lei, Ponman, T. J., Raychaudhury, S., Dariush, A., 2012, ApJ, 757, 122
	\bibitem[\protect\citeauthoryear{Rasmussen, Ponman, Mulchaey}{Rasmussen et al.}{2006}]{ras06} Rasmussen, J., Ponman, T. J., Mulchaey, J. S., 2006, MNRAS, 370, 453
	\bibitem[Rines et al.(2016)]{rin16} Rines, K. J., Gellar, M. J., Diaferio, A., Hwang, H. S., 2016, MNRAS, 819, 63
	\bibitem[Salim et al.(2007)]{sal07} Salim, S., et al., 2007, ApJS, 173, 267
	\bibitem[Schiminovich et al.(2010)]{sch10} Schiminovich, D., et al. 2010, MNRAS, 408, 919
	\bibitem[Searle \& Sargent(1972)]{sea72} Searle, L. \& Sargent, W. L. W. 1972, ApJ, 173, 25
	\bibitem[Serra et al.(2011)]{ser11} Serra, A. L., Diaferio, A., Murante, G., Borgani, S., 2011, MNRAS, 412, 800
	\bibitem[Shimakawa et al.(2015)]{shi15} Shimakawa, R., Kodama, T., Tadaki, K., Hayashi, M., Koyama, Y., Tanaka, I. 2015, MNRAS, 448, 666
	\bibitem[Skillman et al.(1996)]{ski96} Skillman, E. D., Kennicutt, R. C., Jr., Shields, G. A., Zaritsky, D., 1996, ApJ, 462, 147
	\bibitem[Solanes et al.(2001)]{sol01} Solanes, J. M., Manrique, A., Garc\'{i}a-G\'{o}mez, C. Gonz\'{a}lez-Casado, G., Giovanelli, R., Haynes, M. P. 2001, ApJ, 548, 97
	\bibitem[Spitoni(2015)]{spi15} Spitoni, E. 2015, MNRAS, 451, 1090
	\bibitem[Spitoni et al.(2010)]{spi10} Spitoni, E., Calura, F., Matteucci, F., Recchi, S. 2010, A\&A, 514, 73
	\bibitem[Takada et al.(2014)]{tak14} Takada, M., et al. 2014, PASJ, 66, 1
	\bibitem[Tegmark et al.(2004)]{teg04} Tegmark, M., Blanton, M. R., Strauss, M. A., et al., 2004, ApJ, 606, 702
	\bibitem[Tempel et al.(2011)]{tem11} Tempel, E., Saar, E., Liivam\"{a}gi, L. J., Tamm, A., Einasto, J., Einasto, M., M\"{u}ller, V. 2011, A\&A, 529, 53
	\bibitem[\protect\citeauthoryear{Tempel, Tago, \& Liivam\"{a}gi}{Tempel et al.}{2012}]{tem12} Temepl, E., Tago, E., Liivam\"{a}gi, L. J., 2012, A\&A, 540, 106
	\bibitem[\protect\citeauthoryear{Thomas, Greggio, \& Bender}{Thomas et al.}{1998}]{tho98} Thomas, D., Greggio, L., Bender, R., 1998, MNRAS, 296, 119
	\bibitem[Torrey et al.(2012)]{tor12} Torrey, P., Cox, T. J., Kewley, L., Hernquist, L., 2012, ApJ, 746, 108
	\bibitem[Tran et al.(2015)]{tra15} Tran K.-V., et al. 2015, ApJ, 811, 28
	\bibitem[Tremonti et al.(2004)]{tre04} Tremonti, C. A., et al. 2004, ApJ, 898, 913
	\bibitem[Valentino et al.(2015)]{val15} Valentino, F., et al. 2015, ApJ, 801, 132
	\bibitem[Wang et al.(2016)]{wan16} Wang, J., Koribalski, B. S., Serra, P., van der Hulst, T., Roychowdhury, S., Kamphuis, P., Chengalur, J. N., 2016, MNRAS, 460, 2143
	\bibitem[Zahid et al.(2012)]{zah12} Zahid, H. J., Dima, G. I., Kewley, L. J., Erb, D. K., Dav\"{e} R. 2012, ApJ, 757, 54
	\bibitem[Zahid et al.(2013)]{zah13} Zahid, H. J., Geller, M., Kewley, L. J., Hwang, H. S., Fabricant, D. G., Kurtz, M. J. 2013, ApJL, 771, 19
	\bibitem[\protect\citeauthoryear{Zahid, Kewley, \& Bresolin}{Zahid et al. }{2011}]{zah11} Zahid, H. J., Kewley, L. J., Bresolin, F., 2011, ApJ, 730, 137
	\bibitem[Zahid et al.(2014a)]{zah14a} Zahid, H. J., Torrey, P., Vogelsberger, M., Hernquist, L., Kewley, L., Dav\'{e}, R. 2014, Ap\&SS, 349, 873
	\bibitem[Zahid et al.(2014b)]{zah14b} Zahid, H. J., et al., 2014, ApJ, 791, 130
	\bibitem[Zhang et al.(2009)]{zha09} Zhang, W., Li, C., Kauffmann, G., Zou, H., Catinella, B., Shen, S., Guo, Q., Chang, R., 2009, MNRAS, 397, 1243
\end{thebibliography}
\end{document}